\documentclass[referee]{sn-jnl}
\usepackage[utf8]{inputenc}
\usepackage{amsmath}
\usepackage[T1]{fontenc}
\usepackage{caption}
\usepackage{subcaption}
\usepackage{float}
\usepackage{graphicx}
\usepackage{adjustbox}
\usepackage{hyperref}
\usepackage{xr-hyper}
\usepackage{nameref}

\usepackage{xcolor,colortbl}
\PassOptionsToPackage{usenames,dvipsnames}{xcolor}
 
\usepackage{tikz}
\usetikzlibrary{arrows.meta}
\usetikzlibrary{positioning}

\makeatletter
\newcommand*{\addFileDependency}[1]{
  \typeout{(#1)}
  \@addtofilelist{#1}
  \IfFileExists{#1}{}{\typeout{No file #1.}}
}
\makeatother

\newcommand*{\myexternaldocument}[1]{%
    \externaldocument{#1}%
    \addFileDependency{#1.tex}%
    \addFileDependency{#1.aux}%
}

\Newlabel{1}{1}
\Newlabel{2}{2}
\Newlabel{3}{3}
\Newlabel{4}{4}
\Newlabel{5}{5}

\myexternaldocument{supp}

\begin{document}

\title{Pedestrian mobility citizen science complements expert mapping for enhancing inclusive neighborhood placemaking}

\author[1,2,*]{Ferran Larroya}
\author[3]{Roger Paez}
\author[3]{Manuela Valtchanova}
\author[1,2,**]{Josep Perell\'o}
\affil[1]{OpenSystems Research Group, Departament de F\'isica de la Mat\`eria Condensada, Universitat de Barcelona, Mart\'i i Franqu\`es, 1, 08028, Barcelona, Catalonia, Spain}
\affil[2]{Universitat de Barcelona Institute of Complex Systems, Mart\'i i Franqu\`es, 1, 08028, Barcelona, Catalonia, Spain}
\affil[3]{ELISAVA Barcelona School of Design and Engineering, Universitat de Vic- Universitat Central de Catalunya, La Rambla, 30-32, 08002 Barcelona, Catalonia, Spain}
\affil[*]{ferran.larroya@ub.edu}
\affil[**]{josep.perello@ub.edu}

\abstract{Cities are complex systems that demand integrated approaches, with increasing attention focused on the neighborhood level. This study examines the interplay between expert-based mapping and citizen science in the {\it Primer de Maig} neighborhood of Granollers, Catalonia, Spain—an area marked by poor-quality public spaces and long-standing socio-economic challenges. Seventy-two residents were organized into 19 groups to record their pedestrian mobility while engaging in protocolized playful social actions. Their GPS identified opportunity units for meaningful public space activation. Although 56\% of observed actions occurred within expert-defined units, the remaining 44\% took place elsewhere. Clustering analysis of geo-located action stops revealed seven distinct clusters, highlighting overlooked areas with significant social potential. These findings underscore the complementarity of top-down and bottom-up approaches, demonstrating how citizen science and community science approaches enriches urban diagnostics by integrating subjective, community-based perspectives in public space placemaking and informing inclusive, adaptive sustainable urban transformation strategies.}

\maketitle

\section*{Introduction}

Cities function as complex systems \cite{Batty_2013, Barthelemy_2016, Bettencourt_2021}, where multiple dimensions—physical, economic, social, cultural, and emotional—intersect and overlap \cite{Gravier_2024, Verbavatz_2020, Henderson_2007, Barthelemy_2013, Bettencourt_2007,Sennett_2020}. In urban contexts, the neighborhood scale has gained attention, as highlighted in a recent United Nations Habitat report \cite{UN_Habitat_2023}. Neighborhoods represent spatially defined units with distinct functional and social networks, offering an enabling environment for enhancing residents' quality of life. Daily life and social interactions are shaped by the physical characteristics, distribution, and ambiance of public spaces such as streets, squares, and parks \cite{Jackson_2003, Mazumdar_2018, Carmona_2019, Sonta_2023}. Achieving an integrated approach to urban analysis requires consideration of diverse aspects—including social, economic, environmental, and mobility factors—as well as multiple urban dimensions, such as open public spaces and building units \cite{Sennett_2020, UN_Habitat_2023, Carmona_2019}.

The significance of involving existing communities in urban sustainability and related transformative initiatives—particularly at the neighborhood level \cite{Parker_2023}—has been increasingly recognized in both academic research \cite{Healey_1998, Gehl_2013,Rodela_2025,Szaboova_2024} and grassroots movements \cite{pps_web}. This principle is also emphasized in recent policy reports addressing urban societal challenges, such as the EPA’s Climate Change and Social Vulnerability in the United States (2021) \cite{EPA_2021} and the European Commission Sustainable Urban Development Strategies Handbook (2020) \cite{EU_2020}. In this context, placemaking seeks to address residents' needs while reimagining the potential of parks, downtowns, waterfronts, plazas, neighborhoods, streets, markets, campuses, and public buildings \cite{pps_web,Thomas_2016,Ellery_2021}. It encompasses a range of strategies that can eventually integrate participatory practices and multidisciplinary perspectives \cite{Thomas_2016,Ellery_2021}.

Community-driven placemaking is gaining momentum in cities worldwide \cite{Palmer_2024, Sandercock_2024}. It serves as a means to reclaim public spaces and strengthen the sense of place \cite{Toolis_2017}, reinforcing connections between people and their shared environments. Placemaking promotes inclusivity \cite{Omholt_2019} and contributes to the creation of more equitable cities \cite{Couper_2023}. These efforts help amplify the voices of underserved populations and enhance gender \cite{Lorono_2023} and age \cite{Sutton_2002, Lager_2021} perspectives while prioritizing the needs of the most vulnerable \cite{Toolis_2017, Yildiz_2024}. Some placemaking initiatives, with less or more intensive community involvement, focus on integrating nature into urban spaces to promote environmental sustainability and improve residents’ quality of life \cite{Gulsrud_2018, Boros_2021}. Others emphasize mobility by enhancing walkability \cite{Sonta_2023, Lager_2021} and expanding cycling infrastructure \cite{Ferster_2017}.

Not only in community-driven placemaking but also in the broader field of urban science \cite{Acuto_2018}, mapping is a fundamental tool that relies on datasets with geo-located information. Mapping, as an analytical tool, identifies existing opportunities within a given environment and supports urban transformation efforts \cite{Harley_2002, Sandercock_2023, Paez_2024a, Paez_2024b}. Maps not only can capture physical site conditions but also reveal activation flows, time-based processes, and relational dynamics—critical aspects in human mobility data-driven urban science \cite{Xu_2023}. Such approaches, which heavily rely on digital traces, have been applied to a wide range of topics, from the {\it 15-minute city} concept \cite{Moreno_2021, Bruno_2024, Abbiasov_2024} to studies on urban segregation and inequality \cite{Arvidsson_2023,Xu_2025}. A particularly relevant aspect of placemaking is pedestrian mobility, which can be further analyzed through sidewalk infrastructure \cite{Rhoads_2023b} or digital traces of GPS-recorded pedestrian trajectories \cite{Helbing_2001, Jiang_2016, Bongiorno_2021, Hunter_2021}. However, micro-mobility and pedestrian data are often not openly accessible \cite{Yabe_2024}, which poses challenges for walkability studies and urban planning at the neighborhood scale. As a result, the placemaking approach cannot fully leverage this information.

While some mobile phone-based companies release mobility datasets \cite{Yabe_2024b}, citizen science has the potential to generate open pedestrian data \cite{Perello_2024,Larroya_2023a}. Mapping is also a key tool in citizen science, enabling the generation of crowdsourced knowledge and aggregating individual geolocated data \cite{Haklay_2012}. Citizen science involves the active engagement of the general public in scientific research and participation may range from formulating research questions and developing experimental protocols to collecting and interpreting data \cite{Vohland_2021}. A broad interpretation of citizen science is adopted here \cite{Haklay_2021}, encompassing related terms such as community science and participatory research, while remaining attentive to the nuanced subtleties and ongoing debates surrounding these concepts \cite{Cooper_2021}. 

Beyond terminological nuances, citizen science has witnessed significant growth in urban contexts in recent years \cite{Bonhoure_2024}. Public citizen science experiments can be designed to address specific urban issues \cite{Perello_2024, Sagarra_2016, Perello_2022}. Mobility-focused citizen science projects vary in scope but in all cases they can enrich urban analysis by capturing geo-located features such as stops or velocity patterns \cite{Barbosa_2018}. Some projects monitor street traffic with minimal citizen involvement \cite{Janez_2022}, others produce outputs for policymakers \cite{Keseru_2019}, and some specifically engage cyclists \cite{Ferster_2017, Pappers_2022} or pedestrians navigating a science festival placed in a public park \cite{Gutierrez_2016}. Studies on pedestrian behavior have also incorporated gamification strategies to assess walkability \cite{Kapenekakis_2017}, with particular attention to communities facing mobility challenges, such as older adults \cite{Ertz_2021} or adolescents \cite{Larroya_2023a}. Data-driven participatory initiatives \cite{Mueller_2018, Christine_2021} hold significant potential \cite{Toomey_2020}, particularly when they incorporate local community perspectives and address pressing urban social issues \cite{Ramirez_2014, Croese_2021, Grootjans_2022}. These efforts might be seen  within the framework of {\it citizen social science} \cite{Albert_2021, Bonhoure_2023} that starts participatory scientific research from communities' shared social concern and thus align closely with the core goals of placemaking (e.g., enhancing residents' well-being and fostering inclusive, equitable neighborhoods \cite{Toolis_2017}).

This study wants to contribute to the advancement of a citizen science that expands urban visions and experiential mapping within the placemaking framework \cite{Pitidis_2024}. We therefore designed a pedestrian mobility citizen science experiment in the neighborhood of {\it Primer de Maig} in Granollers, Catalonia, Spain, to explore how public space can be better utilized and activated for broader community engagement using citizen science data \cite{Perello_2024,Larroya_2023a,Larroya_2024} and data science techniques such as clustering \cite{Hartigan_1979,Bholowalia_2014,Rousseeuw_1987,Tibshirani_2001}. Built in 1957, {\it Primer de Maig} is one of the city’s oldest modern housing developments and has 2\,964 inhabitants. It is densely populated, covering approximately 3 hectares, with a population density of about 45\,000 inhabitants per ${\rm km}^2$. The neighborhood features a distinct urban structure characterized by pedestrian streets and open spaces between mid-rise buildings, yet it suffers from poor public space quality. {\it Primer de Maig} has historically faced socio-economic challenges, including lower income levels. These economic constraints have contributed to issues such as the deterioration of public spaces and limited community resources.

72 residents were organized into 19 groups, each with well-defined and homogeneous socio-demographic profiles (e.g., children, educators, families, and older adults). The subjective perceptions, narrative landscapes, and symbolic imaginaries can be concretely captured with a citizen science experiment through geolocated pedestrian data as walkability is deeply linked with the concept of active living environment \cite{Ravazzoli_2017,Tobin_2022}. The experiment reported here considers playful and festive activities \cite{Paez_2024c} (we will call them actions), to help inspiring communities to collectively reimagine and reinvent their public spaces \cite{Harley_2002, Sandercock_2023,Pitidis_2024}. Each group was tracked with digital devices as they independently explored the neighborhood looking for most suitable places for running a set of group actions. During the experiment, the groups engaged in the same festive and playful social actions in public spaces, with each group determining the location and duration of their own actions. The collected GPS data underwent a filtering and processing procedure detailed in a dedicated Data Descriptor publication \cite{Larroya_2024}, that also reports the co-design effort with local organizations and residents, and some of the impacts in a municipal level. 

The citizen science pedestrian mobility experiment was conducted in parallel with an independent cartographic study of the neighborhood led by urban experts. By combining and contrasting both approaches, this research aims to offer a more comprehensive understanding on how citizen science can contribute to urban sustainability and to more inclusive and participatory urban placemaking and urban transformation. Ultimately, this study underscores the importance of incorporating personal, lived experiences into urban explorations in a neighborhood.

\section*{Results}

\subsection*{Cartographic expert-based mapping}

The {\it Primer de Maig} neighborhood comprises 41 buildings, housing a total of 550 residential units and 12 commercial spaces. It has a total area of 26\,429 m$^2$, of which 11\,774 m$^2$ (44.5\%) are designated as pedestrian spaces. An expert-based analysis conducted an in-depth study of the neighborhood's urban structures, generating maps that identify areas with high potential for appropriability from visual, spatial, and social perspectives \cite{Paez_2024b}. The aim of these initial mappings was to reveal areas of opportunity where placemaking initiatives could flourish. This was an important step to characterize the differences in a seemingly homogeneous urban fabric. Beyond its analytical capacities, the approach favors using mapping as a tool to inquire on the propensities and potentialities of urban space. This projective agency frames mapping as a design tool, allowing for a more nuanced approach to urban regeneration processes, as well as a more precise, strategic allocation of efforts and resources. Specifically, experts-based approach mapped visual, spatial, and social features of the neighborhood (see Methods).

\begin{figure}[h]
\centering
\includegraphics[width=0.95\linewidth]{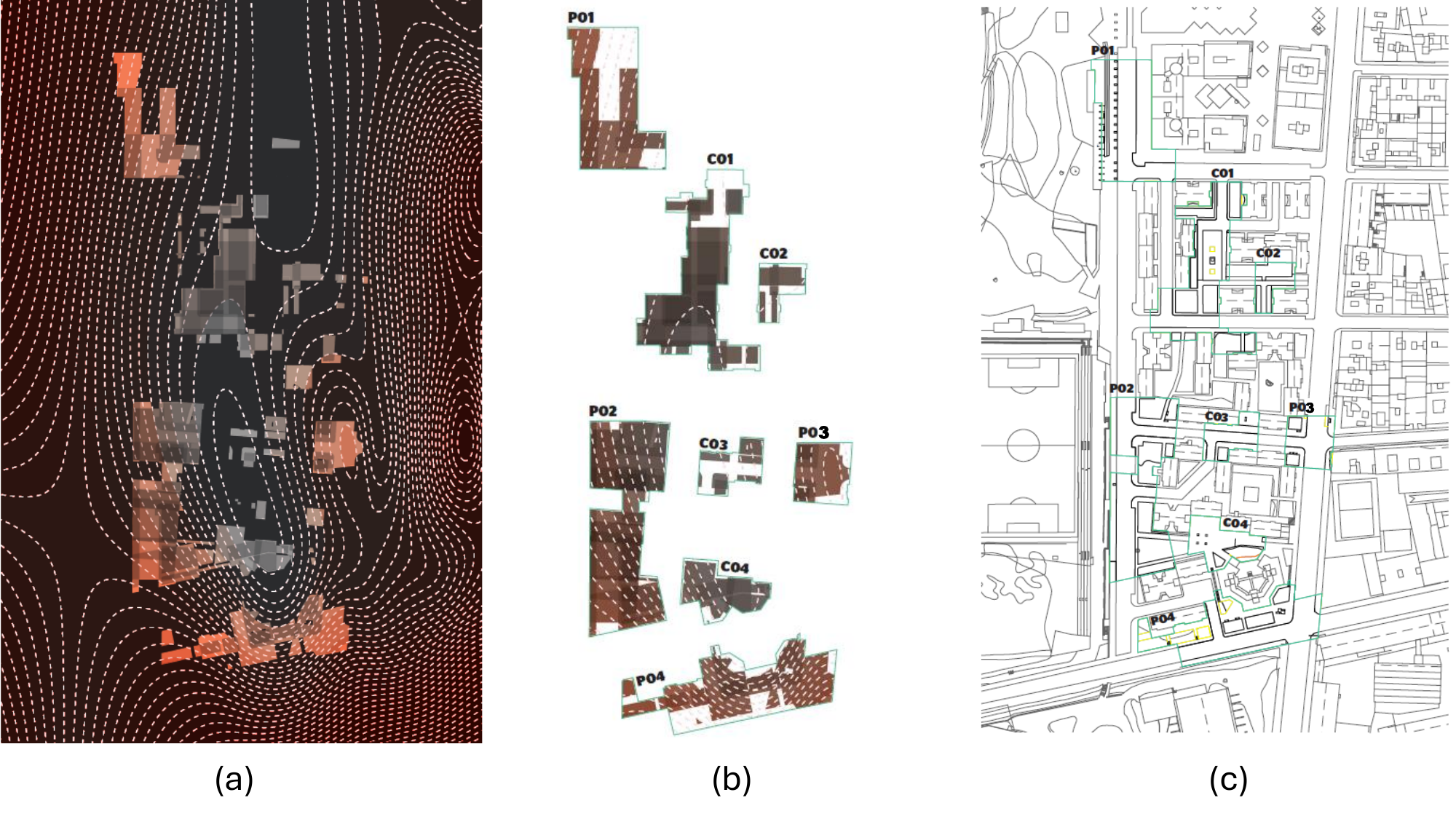}
\caption[Identification of Opportunity Units with cartographic expert-based mapping.]{\textbf{Identification of opportunity units with cartographic expert-based mapping.} (a)  Intersection of the three maps—the visual, spatial, and social dimensions—revealing eight areas with high appropriability potential by the residents. Darkest areas are the most promising ones. (b) The eight areas of appropiability (opportunity units, OUs) with their code-name. C01, C02, C03 and C04 correspond to inner areas of the neighborhood while P01, P02, P03 and P04 operate as portals, due to their capacity to redistribute pedestrian flow the neighborhood. (c) Map of the neighborhood in white with the eight OUs. The {\it Primer de Maig neighborhood} in Granollers, Catalonia, Spain, covers an area of 3 hectares (300 meters by 100 meters approximately).}
\label{fig:uoa2}
\end{figure}

The intersection of three mapping dimensions—visual, spatial, and social—reveals eight areas with high potential for public space activation (see Figure \ref{fig:uoa2}a). These areas, identified as opportunity units (OUs), combine visual prominence, spatial approapriability, and socio-cultural activity. The darkest regions on the Figure \ref{fig:uoa2}b map indicate the highest activation potential, with the most prominent located in the neighborhood's inner sections, particularly in OU C01. Additionally, some areas function as key entry points, redistributing pedestrian flow—most notably, P02. Figure \ref{fig:uoa2}c provides a broader context by incorporating streets and building units both within and surrounding the neighborhood. The map series provided here and in the Supplementary Information reveals opportunities hitherto unknown, and thus opens up new possibilities to transform the urban setting, and, specifically, informs the participatory process and the citizen science experiment.

\subsection*{Pedestrian exploratory mobility patterns in the citizen science experiment}\label{ss:basic_statistics}

\begin{figure}[h]
\centering
\includegraphics[width=0.9\linewidth]{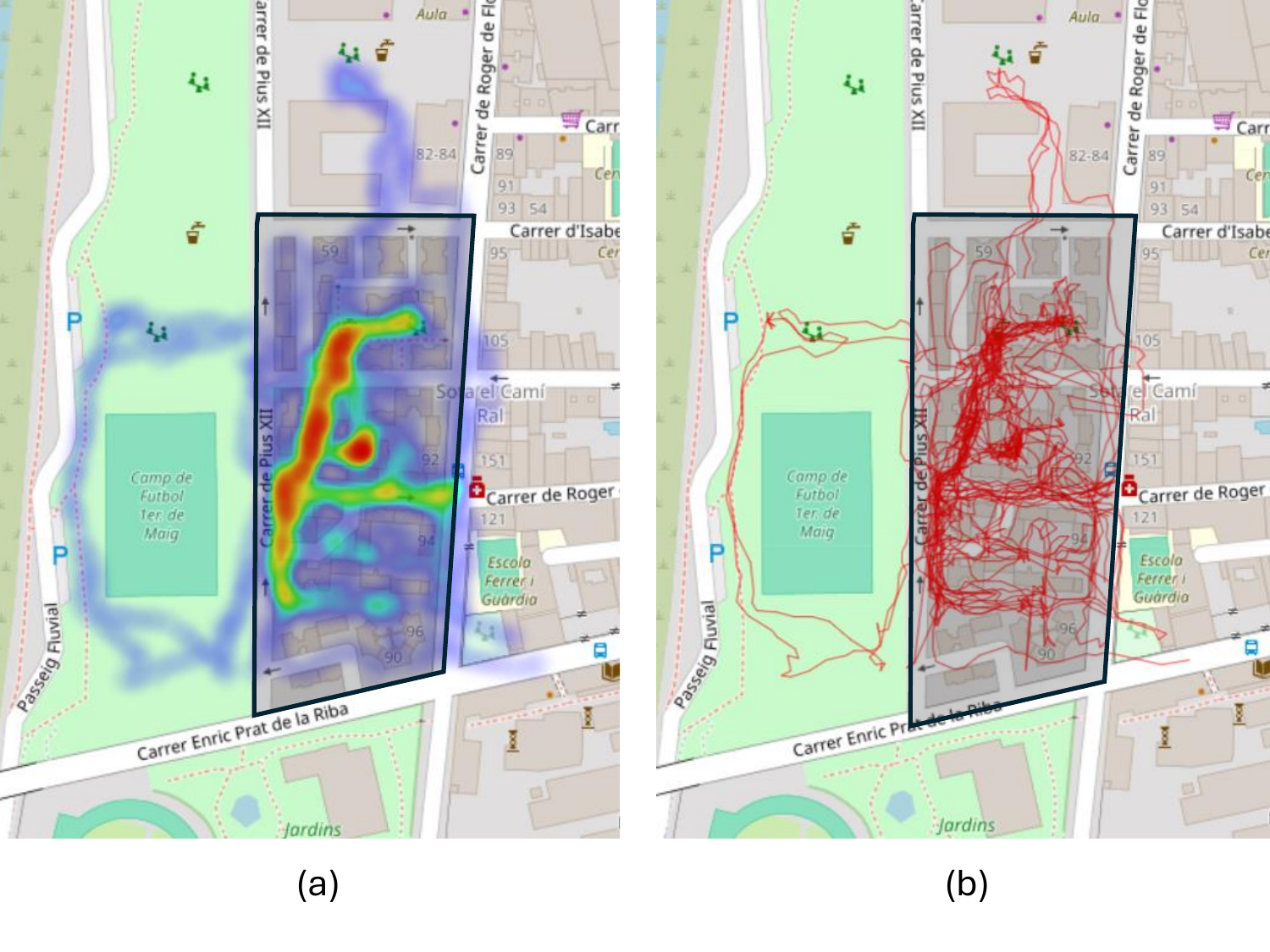}
\caption{{\bf Spatial distribution and trajectories of participant groups in the citizen science exploratory pedestrian mobility experiment}. (a) Heatmap of all recorded GPS positions. (b) Trajectories of the 19 participant groups. In both visuals, the black perimeter outlines the {\it Primer de Maig} neighborhood. The map is rendered using the OpenStreetMap layout.}
\label{fig:map}
\end{figure}

The group participants of the citizen science experiment were asked to complete six social actions of festive and playful nature (sitting around chatting, playing ball, dancing to music, snacking, toasting and shooting confetti). Full details of the data collected --jointly with the experiment protocols and design-- are provided in the Data Descriptor of this manuscript \cite{Larroya_2024}. During the experiment, which is also described in Methods, the groups of participants independently chose most suitable public spaces for performing the actions while they walk around the neighborhood. Figure \ref{fig:map}a provides a heatmap visualization of aggregated exploratory pedestrian data, illustrating the density of GPS records throughout the experiment. The 19 trajectories recorded by the participating groups are shown in Figure \ref{fig:map}b.

The collected pedestrian mobility data reflects relatively short trajectories in both time and distance (see Table \ref{tab:basic_statistics}). On average, participant groups spent approximately 1 hour and 10 minutes ($T = 4\,199 \pm 352$ s, see Eq. (\ref{eq:T}) in Methods) completing the six actions while exploring the neighborhood, within the allotted 1 hour and 30 minutes. As shown in Table \ref{tab:basic_statistics}, the average total distance covered was $D = 955 \pm 92$ m (cf. Eq. (\ref{eq:D}), see Methods), which is relatively short compared to the total duration of their trajectories, as participants spent a significant portion of their time stationary while freely interpreting and performing designated actions. This is further reflected in the low average effective speed (total distance covered divided by trajectory duration), measured at $v_{\rm eff}=D/T= 0.24 \pm 0.02$ m/s ($0.86 \pm 0.07$ km/h). When in motion, however, participants moved at an average instantaneous velocity of $v=1.38 \pm 0.03$ m/s (cf. Eq. (\ref{eq:v}), see Methods). These velocities are significantly smaller compared to those obtained in another citizen science experiment, where participants had predefined origin-destination goals \cite{Perello_2024,Larroya_2023a}. Table \ref{tab:basic_statistics} presents additional statistical metrics (quartiles, minimum, and maximum values) for trajectory duration, distance covered, effective speed, and instantaneous velocity.

Among the different features that characterize pedestrian mobility \cite{Barbosa_2018}, the stops are the most relevant to provide insights on the placemaking and appropriability potential of each of the public space sites of the neighborhood. The stops were identified by analyzing the time difference between consecutive timestamps of the collected GPS records, using a threshold of 10 seconds as the minimum time difference to consider a record as ``stop'' (see Methods). The total number of stops detected was 339. Several statistical analysis can be performed. On average, each stop lasted about 3 minutes ($\tau_{\rm All}=191 \pm 20$ s) but it is of particular interest here to focus on those stops in which groups performed the actions. The stops corresponding to the actions were identified extracting the GPS location of the videos recorded by the participants when performing the actions (see Methods). 

When considering only the 81 stops specifically dedicated to the actions, the average stop duration increased to over 12 minutes ($\tau_{\rm Actions}=758 \pm 73$ s). As shown in Table \ref{tab:basic_statistics}, half of the actions (interval Q1-Q3) lasted between 5 and 16 minutes. There are some punctual actions of very few minutes of duration (39 seconds the shortest one) and others of much longer duration (more than 30 and 40 minutes) due to the fact that some of the actions were carried out simultaneously/sequentially by some of the groups on the same location. We also detected 13 stops took place outside the neighborhood, in the area of the park and municipal soccer pitch (left side of the map in Figure \ref{fig:map}). These stops were excluded from the analysis in the forthcoming steps.

\begin{table}[t]
\centering
\begin{adjustbox}{width=\textwidth}
\begin{tabular}{lrrrrrrr}
\hline \hline
& $\langle \dots \rangle$ & $\sigma$ & Q1 ($25\%$) & Q2 ($50\%$) & Q3 ($75\%$) & min & max \\ \hline
$D$ (m) & $955\pm 92$ & $389$ & $768$ & $883$ & $1\,257$ & $280$ & $1\,815$ \\
$T$ (s) & $4\,199\pm 352$ & $1\,493$ & $3\,746$ & $4\,045$ & $5\,063$ & $694$ & $7\,407$ \\
$v_{\rm eff}$ (m/s) & $0.24\pm 0.02$ & $0.09$ & $0.18$ & $0.24$ & $0.32$ & $0.09$ & $0.40$ \\
$v$ (m/s) & $1.38\pm 0.03$ & $1.59$ & $0.89$ & $1.14$ & $1.41$ & $0.56$ & $30.38$ \\
$\tau_{\rm All}$ (s) & $191\pm 20$ & $370$ & $13$ & $36$ & $199$ & $10$ & $2\,669$ \\
$\tau_{\rm Actions}$ (s) & $758\pm 73$ & $657$ & $308$ & $525$ & $963$ & $39$ & $3\,130$ \\
\hline\hline
\end{tabular}
\end{adjustbox}
\caption{\label{tab:basic_statistics} {\bf Main statistics of the trajectories in the pedestrian mobility citizen science experiment.} Mean value with the standard error of the mean, standard deviation $\sigma$, quartiles (Q1, Q2, and Q3) and minimum (min) and maximum (max) values of the total distance $D$ covered, the trajectory duration $T$, the effective speed $v_{\rm eff}=D/T$, the instantaneous velocity $v$ and the stops durations ($\tau_{\rm All}$ and $\tau_{\rm Actions}$). A total of 339 stops were identified (All) and 81 stops were attributed to the actions (Actions).}
\end{table}

\subsection*{Spatial analysis of citizen science data}

\begin{table}[t]
\centering
\begin{adjustbox}{width=\textwidth}
\begin{tabular}{lrrrrrr}
\hline \hline
OUs & stops (\%) & $\sum\tau_i$ (s) & $\langle \tau \rangle$ (s) & $\sigma_{\tau}$ (s) & $\tau^{\rm min}$ (s) & $\tau^{\rm max}$ (s) \\ \hline
P01 & $0$ $(0\%)$ & $0$ $(0\%)$ & $0$ & $0$ & $0$ & $0$\\
P02 & $17$ $(25\%)$ & $12\,739$ $(24\%)$ & $749\pm 126$ & $520$ & $100$ & $1\,955$ \\
P03 & $0$ $(0\%)$ & $0$ $(0\%)$ & $0$ & $0$ & $0$ & $0$ \\
P04 & $0$ $(0\%)$ & $0$ $(0\%)$ & $0$ & $0$ & $0$ & $0$\\
C01 & $9$ $(13\%)$ & $9\,904$ $(19\%)$ & $1\,100\pm 273$ & $818$ & $96$ & $2\,677$\\
C02 & $7$ $(10\%)$ & $3\,517$ $(7\%)$ & $502\pm 109$ & $290$ & $100$ & $909$ \\
C03 & $1$ $(1\%)$ & $1\,845$ $(3\%)$ & $1,845$ & $0$ & $1\,845$ & $1\,845$ \\ 
C04 & $4$ $(6\%)$ & $3\,805$ $(7\%)$ & $951\pm 507$ & $1\,014$ & $39$ & $2\,669$  \\ \hline
All OUs & $38$ $(56\%)$ & $31\,810$ $(60\%)$ & $837\pm 111$ & $687$ & $39$ & $2677$
\\
All actions & $68$ $(100\%)$ & $53\,060$ $(100\%)$& $780\pm 83$ & $680$ & $39$ & $3\,130$ \\
\hline\hline
\end{tabular}
\end{adjustbox}
\caption{\label{tab:action_units} {\bf Action stops within the opportunity units.} Number of stops, total duration of stops, average stop duration with the standard error of the mean, standard deviation and the shortest and the longest stop for each one of the eight opportunity units (OUs). The last two rows represents the statistics within the neighborhood (68 stops) and within the eight OUs (38 stops).}
\end{table}

We next combined and contrasted the expert-based mapping with findings from the exploratory pedestrian mobility citizen science experiment. We analyzed the 68 stops that can be attributed to the actions and that are located within the neighborhood. 38 of these actions (56\%) took place within the OUs, while 30 (44\%) occurred elsewhere. Notably, this result is consistent with the fact that the sum of the stops duration within the OUs accounted for only 60\% of the sum of all stops duration for performing the six group actions ($\sum_{i\in OU}\tau_i=31\,810$ seconds from the 38 stops within the OUs versus $\sum\tau_i=53\,060$ seconds taking all 68 stops within the neighborhood). 

\begin{figure}[ht]
\centering
\includegraphics[width=0.9\linewidth]{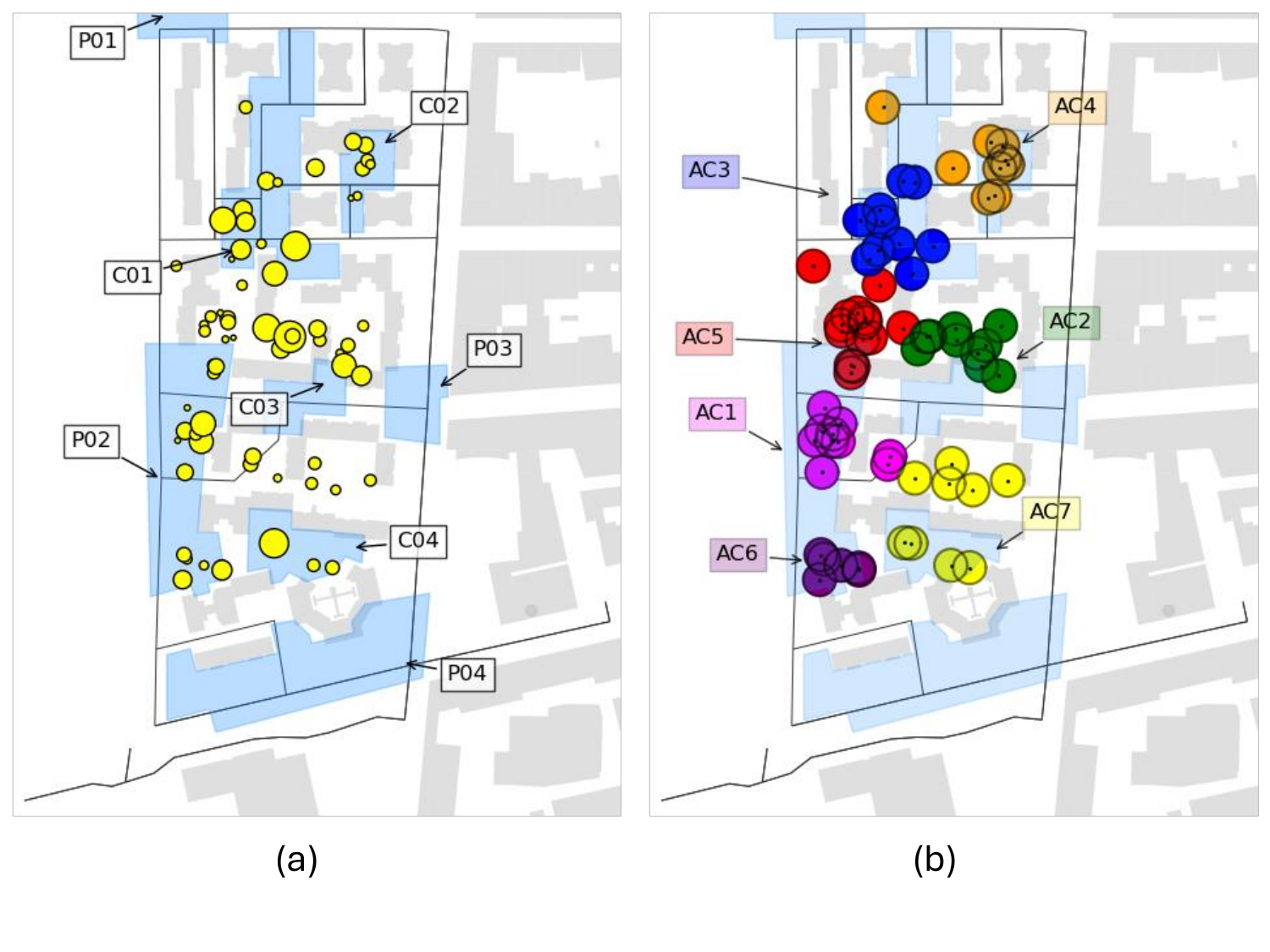}
\caption{{\bf Location of the 68 action stops within the neighborhood.} (a) The eight opportunity units (OUs) independently defined by the urban experts correspond to the blue areas. The yellow dots are the locations of the 68 stops. The bigger they are, the longer the stop duration. (b) The seven clusters of stops applying the weighted $K$-means algorithm. For visualization purposes, all stops are represented with bubbles of the same size. Each cluster uses a different color. The map is rendered using the OpenStreetMap layout.}
\label{fig:pausing_events}
\end{figure}

Table \ref{tab:action_units} provides a detailed breakdown of the number and duration of the stops taking place within the OUs, highlighting variations in their alignment with expert-defined areas evaluation in terms of their importance (cf. Figure \ref{fig:uoa2}). Most of the OUs categorized as portals do not include any stop, except for one (P02) which includes a relevant number of stops (17, 25\% over the total number of action stops). This portal unit (P02) is especially attractive due to its large composition of green flowerbeds, adjacent to the street that links the neighborhood with the green areas of the park nearby ({\it Parc del Congost}). In the middle of the unit, there is also a pedestrian walkway connecting two streets. Concerning the OUs located in the inner areas of the neighborhood, the C01 unit is the one containing the larger number of action stops and largest average stop duration, and the largest stop. The result is consistent with the urban expert-based mapping which identifies C01 as the area with highest appropriability potential. Figure \ref{fig:pausing_events}a complements the analysis by displaying a neighborhood map, where opportunity units (OUs) are highlighted in blue, and stops are represented by yellow dots, sized according to their duration.

Figure \ref{fig:map}b illustrates that the trajectories extensively cover the neighborhood. In contrast, the heatmap in Figure \ref{fig:map}a reveals significant heterogeneity in the distribution of GPS data, with concentrations primarily near or within the P02 and C01 units. Additionally, Figure \ref{fig:pausing_events}a demonstrates that stops are unevenly distributed. This uneven distribution has led us to employ more advanced techniques to directly identify areas of interest using citizen science data. We have taken the simplest possible data clustering analysis. We apply the well-known $K$-means algorithm \cite{Hartigan_1979} to the 68 pausing events, in its weighted version and using the stop duration as weight. The Methods section and Supplementary Material offer detailed technical information about the algorithm employed and discuss alternative approaches that yielded less successful outcomes. We place particular emphasis on the selection of the optimal number of clusters, referencing established methodologies \cite{Bholowalia_2014,Rousseeuw_1987,Tibshirani_2001}. After conducting multiple analyses, we determined that setting $K=7$ clusters was most appropriate. Figure \ref{fig:pausing_events}b presents the clustering results, while Table \ref{tab:clusters_weighted_kmeans} provides comprehensive statistics for each cluster, including the number of stops, total stop duration, average stop duration, standard deviation, and the shortest and longest stop durations.

\begin{table}[t]
\centering
\begin{adjustbox}{width=\textwidth}
\begin{tabular}{lrrrrrr}
\hline \hline
Cluster & stops (\%) & $\sum \tau_i$ (s) & $\langle \tau \rangle$ (s) & $\sigma_{\tau}$ (s) & $\tau^{\rm min}$ (s) & $\tau^{\rm max}$ (s) \\ \hline
AC1 (magenta) & $10$ $(15\%)$ & $8,083$ $(15\%)$ & $808\pm 202$ & $638$ & $100$ & $1\,955$\\
AC2 (green) & $10$ $(15\%)$ & $10\,532$ $(20\%)$ & $1\,053\pm 276$ & $872$ & $200$ & $3\,130$ \\
AC3 (blue) & $10$ $(15\%)$ & $11\,786$ $(22\%)$ & $1\,179\pm 270$ & $854$ & $96$ & $2\,677$ \\
AC4 (orange) & $9$ $(13\%)$ & $4\,999$ $(9\%)$ & $555\pm 104$ & $311$ & $100$ & $963$ \\
AC5 (red) & $14$ $(20\%)$ & $7\,761$ $(15\%)$ & $547\pm 150$ & $563$ & $87$ & $2\,345$ \\
AC6 (purple) & $6$ $(9\%)$ & $4\,349$ $(8\%)$ & $725\pm 162$ & $397$ & $262$ & $1\,243$ \\
AC7 (yellow) & $9$ $(13\%)$ & $5\,650$ $(11\%)$ & $628\pm 262$ & $785$ & $39$ & $2\,669$ \\ \hline
All actions & $68$ $(100\%)$ & $53\,060$ $(100\%)$ & $780\pm 83$ & $680$ & $39$ & $3\,130$ \\
\hline\hline
\end{tabular}
\end{adjustbox}
\caption{\label{tab:clusters_weighted_kmeans} {\bf Action stops within the 7 $K$-means weighted clusters.} Number of stops, total stopping time, average stop duration with the standard error of the mean, standard deviation and the shortest and the longest stop for each one of the 7 $K$-means clusters, weighted with the stop duration (see Figure \ref{fig:pausing_events}b).}
\end{table}

The seven clusters (AC1–AC7) identified reflect varied spatial patterns of public space appropriation within the neighborhood. A detailed description of each cluster—regarding stop duration, urban context, and type of actions—is provided in the Supplementary Material (see also Table \ref{tab:actions_clusters}). We highlight the attractiveness of green flowerbeds, which emerge as the predominant urban element across all clusters and the majority of action stops—for instance, in AC5 (red), which exhibits the highest density of stops (14 stops, corresponding to 24 actions). We also emphasize cluster AC4 (orange), centered around a playground and {\it Pla\c ca de la Font}—the neighborhood's only square and a common post-activity gathering spot. However, most actions in {\it Pla\c ca de la Font} took place along its periphery, likely due to the square's hard pavement. AC2 (green), concentrated in an interstitial space featuring a drinking fountain, includes some of the longest action durations and stands out as a socially active node, despite not being identified in the expert-based mapping.

\begin{table}[t]
\centering
\begin{adjustbox}{width=\textwidth}
\begin{tabular}{cccccccccc}
\hline \hline
& P01 & P02 & P03 & P04 & C01 & C02 & C03 & C04 & All \\ \hline
AC1 & 0 & \cellcolor{blue!40}8 & 0 & 0 & 0 & 0 & 0 & 0 & 8  \\
AC2 & 0 & 0 & 0 & 0 & 0 & 0 & \cellcolor{blue!7}1 & 0 & 1  \\
AC3 & 0 & 0 & 0 & 0 & \cellcolor{blue!50}9 & 0 & 0 & 0 & 9  \\
AC4 & 0 & 0 & 0 & 0 & 0 & \cellcolor{blue!30}7 & 0 & 0 & 7 \\
AC5 & 0 & \cellcolor{blue!10}3 & 0 & 0 & 0 & 0 & 0 & 0 & 3 \\
AC6 & 0 & \cellcolor{blue!21}6 & 0 & 0 & 0 & 0 & 0 & 0 & 6  \\ 
AC7 & 0 & 0 & 0 & 0 & 0 & 0 & 0 & \cellcolor{blue!15}4 & 4 \\ 
\hline
All & 0 & 17 & 0 & 0 & 9 & 7 & 1 & 4 & 38 \\
\hline \hline
\end{tabular}
\end{adjustbox}
\caption{\label{tab:cross-stops} {\bf Cross table with the number of action stops between the mapping opportunity units and the 7 detected clusters with $K$-means.} The last row represents, for each opportunity unit (OU, P01-04, C01-04) the total number of stops shared with all clusters (sum of the entire column). Similarly, the last column represent, for a given $K$-means cluster, the sum of all stops shared with all OUs.}
\end{table}

Table \ref{tab:cross-stops} summarizes the distribution of stops across opportunity units (OUs) and clusters identified through $K$-means analysis. While portals P01, P03 and P04 do not collect any stop, P02 collects 17 stops, 8 of which belong to cluster AC1 (out of the 10 in the cluster), all those in cluster AC6 (6 stops) and 3 stops out of the 14 in cluster AC5. This highlights P02's significant role as a central hub, offering diverse areas for various activities due to its expansive size and favorable urban features. Regarding the OUs corresponding to the inner areas of the neighborhood, C01 contains almost all the stops from cluster AC3 (9 out of 10), and C02 most of the AC4 stops (7 of 9). C04 captures the 50\% of the stops from AC7 and C03 only one of the stops from AC2. Notably, many stops from clusters AC5 and AC2 do not fall within any OU, indicating a complex spatial distribution that may not align neatly with predefined boundaries. This analysis underscores the intricate relationship between detected clusters and OUs, suggesting that certain areas attract specific activities, while others transcend these boundaries.

\section*{Discussion}\label{s:discusion}

This work provides a new citizen science based methodology and compares the approach with a cartographic expert-based mapping. To  demonstrate how citizen science enriches urban diagnostics by integrating subjective and community-based perspectives, we have examined pedestrian mobility and placemaking \cite{Thomas_2016,Ellery_2021,Palmer_2024,Sandercock_2024,Toolis_2017} in an urban neighborhood marked by poor-quality public spaces and long-standing socio-economic challenges. 

We take as a reference an expert-based mapping of the {\it Primer de Maig} neighborhood (Granollers, Catalonia, Spain). The expert-based mapping combined spatial data analysis with field observations to capture the complex interplay between the built environment and social dynamics. Maps being produced identify public space areas with high potential for appropriability \cite{Paez_2024b} (opportunity units, OUs): visual aspects considered facades and vertical surfaces as urban landmarks, spatial analysis focused on pedestrian appropriability and movement, and social analysis identified sociocultural attractors.

Given the complexity of urban systems and the importance of neighborhood-level studies in enhancing quality of life \cite{Gehl_2013,Palmer_2024,Sandercock_2024,Couper_2023}, expert-based analyses can be enriched by citizen-led initiatives \cite{Omholt_2019}. In this manuscript, the expert-based mapping serves as a benchmark to highlight how citizen science \cite{Vohland_2021,Haklay_2021} offers distinct contributions. By engaging residents in data collection, citizen science provides alternative or complementary pathways for robust analysis and site diagnostics aimed at fostering positive urban change \cite{Sandercock_2023}. While placemaking aims to create public spaces that foster social interaction \cite{Thomas_2016,Ellery_2021,Palmer_2024,Sandercock_2024,Toolis_2017,Omholt_2019}, citizen science  facilitates knowledge co-production by involving specific groups of concerned citizens—particularly those who are underserved and in vulnerable situations, such as older adults, families, or children—in the research process \cite{Albert_2021, Bonhoure_2023}. Moreover, digital applications for participatory data collection can generate sufficiently large, uniformly formatted datasets to support advanced analyses (e.g., clustering \cite{Hartigan_1979}) and straightforward comparisons with expert-based findings \cite{Ertz_2021,Mueller_2018,Christine_2021}. Additionally, citizen science opens avenues for understanding participants’ emotional responses, thereby creating a more nuanced reading of neighborhood dynamics \cite{Lager_2021} and walkability \cite{Sonta_2023,Tobin_2022} by mapping actions alongside emotional states \cite{Harley_2002}.

Incorporating time-based events and experiences, situated in sociocultural relationships within urban public spaces, pedestrian mobility citizen science experiments \cite{Perello_2024,Larroya_2023a,Larroya_2024} provides an innovative yet rigorous method for analyzing the urban milieu \cite{Bruno_2024,Abbiasov_2024,Arvidsson_2023}. The results highlight the potential of citizen science initiatives to identify additional high-value areas of appropriability beyond those identified through expert-based studies. They also provide detailed insights into the types of actions conducted and their durations.

The {\it Primer de Maig} case study demonstrates how citizen-generated mobility data can complement—or challenge—traditional urban analyses, yielding insights into how public spaces are used for social and festive activities. As said in the first paragraph, {\it Primer de Maig} neighborhood is of particular interest since it suffers a poor quality public space and has historically faced socio-economic challenges. This makes particularly relevant to articulate citizen science initiatives that give the voice to the residents \cite{Albert_2021} in an structured manner, through their pedestrian mobility GPS records \cite{Lager_2021,Helbing_2001}. Importantly, the actions that participants performed were carried out in groups, enabling participants to collectively reflect on the neighborhood’s various public spaces. The citizen science experiment revealed the importance of stationary activities in shaping urban experiences throughout relatively short explorations that are closely related to the {\it 15-minute city}  \cite{Moreno_2021,Bruno_2024,Abbiasov_2024}. Notably, the experiment showed that only 56\% of the recorded actions of festive or social nature took place in the expert-defined OUs, indicating that citizen science can help uncover additional spaces with high social potential that may not be captured by expert-based mappings.

Stop-detection algorithms \cite{Larroya_2024} for pedestrian mobility data were used to identify key areas for public space activation, community appropriation, and potential infrastructure improvements. Spatial clustering of these stopping points deepened the understanding of neighborhood use patterns. A weighted $
K$-means algorithm \cite{Hartigan_1979} identified seven distinct clusters (ACs), each revealing unique patterns of interaction. Some clusters, such as AC3, strongly overlapped with expert-defined areas (e.g., C01), confirming their importance. Others, like AC2, emerged independently, underscoring overlooked spaces with high social value. The diversity of action types within these clusters further suggests that different urban environments foster specific forms of engagement, from playful activities to communal gatherings \cite{Paez_2024a,Paez_2024b}.

These findings underscore the value of participatory methods in urban research, demonstrating that bottom-up, data-driven approaches can effectively complement more conventional urban sustainability planning processes \cite{Ravazzoli_2017}. The pedestrian mobility citizen science experiment took place amid discussions of significant urban transformations in the neighborhood. While the collected data were not central to the planning process, they contributed to ongoing debates. Subsequently, a comprehensive rehabilitation project was launched, involving  41 buildings that house 550 residences and 12 commercial spaces, alongside cultural initiatives aimed at reinforcing neighborhood identity, promoting employment, and revitalizing commercial activity—accompanied by socio-community interventions and improvements to architecture and public spaces. More particularly, there was a renovation of {\it Plaça de la Font}, the area surrounding a drinking fountain that played a central role in cluster AC4, despite the low quality of the public space.

Nonetheless, additional data and more participants within each resident group could yield deeper insights and more robust conclusions. Future work could refine clustering techniques, integrate qualitative feedback from participants, and assess the long-term impact of urban interventions with periodic citizen science pedestrian mobility experiments. Ultimately, merging citizen science with expert urban analysis supports a more integrated understanding of public space usage, informing inclusive and flexible urban design strategies, including temporary space design interventions to foster and support unprogrammed placemaking activities autonomously initiated by neighbors themselves. Citizen science --or community science and participatory research \cite{Haklay_2021,Cooper_2021}-- can thus join forces and become a powerful complementary contribution to those cartographic mappings that seeks to visualize not only geographic data but also the social contexts and relationships within a given area, incorporating lived experiences of communities \cite{Harley_2002,Sandercock_2023,Paez_2024a,Paez_2024b}. These are crucial aspects in the context of urban sustainability \cite{Rodela_2025,Szaboova_2024}.
  
\section*{Methods}

\subsection*{Experiment design}\label{ss:experiment}

The citizen science experiment was co-created together with representatives of three local communities from the neighborhood with well-defined socio-demographics profile to enhance active involvement of and to respond community needs and concerns \cite{Albert_2021}. Co-creation sessions were held with each of the groups to express concerns about the use of public space for social interactions. The sessions also served to develop a robust experimental protocol to collect mobility data that can be used to propose interventions in public space \cite{Senabre_2018,Senabre_2021}. The communities finally involved throughout a larger call for participation were a class from the public school {\it La Gu\`ardia} (11-12 years old) and their teachers, the association of neighbors (with families and thus following a diverse age profile) and a senior citizen's organization (more than 65 years old). The experiment consisted of forming exploration groups of 3-6 people. Each group took an {\it exploration kit}, a bag with different objects to prompt the free performance of 6 specific actions of a festive nature (sitting around chatting, playing ball, dancing to music, snacking, toasting and shooting confetti) which they did not know before starting the experiment \cite{Larroya_2024}. They explored the neighborhood while looking for places to complete the list of actions with a time limit of 1 hour and 30 minutes. Participants walked in small groups through the neighborhood and one participant in each group carried a tablet device with the Wikiloc App \cite{Wikiloc} to record the GPS data of their own trajectory. The use of the Wikiloc App \cite{Wikiloc}, with accounts registered by the researchers and with rented tablet devices, guaranteed the privacy and total anonymity of the participants. We did not collected any personal information on an individual level and all participants signed an informed consent (or their tutors when required). The data collection started and ended in public space, and therefore no additional anonymization method had to be applied (e.g. when the journeys start or end at participants' homes) \cite{Larroya_2023a,Wang_2020}.

\subsection*{Data description}\label{ss:data}
The total amount of processed GPS records is $2\,981$, from 19 pedestrian trajectories corresponding to groups of 3-6 people with different socio-demographic profiles (72 participants in total) \cite{Larroya_2024}. Each group had a unique trajectory described through a collection of GPS coordinates (latitude and longitude) and timestamps. Each trajectory is processed individually, removing GPS locations outside the neighborhood boundaries, at the beginning (and sometimes at the end) of the trajectory. We also obtain the time difference between consecutive timestamps (in seconds), the distance (in metres) and the instantaneous velocity (metres per second) between consecutive GPS records. Each GPS record is labeled as ``moving'' or ``stop'', based on the time difference between consecutive timestamps. Wikiloc App \cite{Wikiloc} that collects the data automatically modulate the timestamps frequency according to the changes in consecutive GPS records. We have chosen a threshold of 10 seconds between consecutive GPS records to consider a record as stop. Therefore, if the following record is collected 10 seconds or more later, the record is labeled as ``stop'' (otherwise, ``moving''). The 10-second threshold is obtained after several validation tests and guarantees that the actual stops made on the route are being captured and that the number of stops is neither over- nor underestimated (see the Data Descriptor \cite{Larroya_2024} for further details). The stops made during the experiment can be due to the actual micro-mobility of the participants (e.g., crossing a street, reorientation processes) but we filter those locations that the participants consciously chose while exploring the neighborhood to carry out the different group actions. In total, we have detected 339 stops, 81 of which correspond to the group actions (although there were 19 groups and 6 different actions to complete, some of them were carried out simultaneously at the same stop). In order to extract the stops corresponding to the actions, we used the locations of the videos recorded by the participants while they were performing the actions. The duration of these is obtained by analyzing the data-set, using the time difference between consecutive timestamps \cite{Larroya_2024}. Lastly, we filtered out only the stops within the neighborhood boundaries, so 13 stops were excluded and the number of stops due to actions was finally 68. A data repository \cite{Larroya_2023b} contains the raw and the processed individual trajectories and stops.

\subsection*{Expert-based mapping}\label{ss:mapping}
The experts-based approach mapped visual, spatial and social features of the neighborhood that identify areas high potential for appropriability.

{\bf Visual: Facades and vertical public space.} A distinctive feature of the neighborhood's structure is the presence of numerous party walls—blank facades without windows—that have substantial street-level visibility. These facades offer the potential to serve as urban landmarks or reference points within the neighborhood’s socio-spatial fabric. In total, 76 facades and over 4\,300 ${\text m}^2$ of vertical surfaces were identified as interacting with the urban space. The visual impact and visibility perimeters of these facades were evaluated. As a result of the mapping process, the urban spaces with significant interaction with these facades were then finally also identified. The final map (Figure \ref{fig:uoa1}a) shows these overlapping areas in varying shades of gray, highlighting the most significant areas of visual interest. The darkest areas (two in Figure \ref{fig:uoa1}a) represent areas where five to seven visibility areas converge.

{\bf Spatial: Degrees of pedestrian appropriability.} The neighborhood's urban fabric was assessed to determine the degree of pedestrian appropriabilty (i.e., how likely is it for the space to be temporarily used by pedestrians beyond mere mobility), assigning four levels based on a grayscale color code as well. Level 0: Private use, shown in white. Level 1: Low  accessibility (vehicle traffic areas), marked with 30\% black. Level 2: Medium accessibility (pedestrian walkways), shown with 50\% black. Level 3: High accessibility (intermediate spaces like green flowerbeds, squares, and pedestrian-only areas), represented with 70\% black. Figure \ref{fig:uoa1}b illustrates the four levels within the neighborhood.

{\bf Social: Attractors of socio-cultural activity.} Public space areas were classified into three main groups, assigning each node of activity with a height in proportion to affluence, i.e., how many people does it attract. Based on the resulting topography, an iso-line map was thus generated, identifying socio-cultural activity around the neighborhood. Education and cultural locations (e.g., schools, libraries, cultural centers) with high affluence, were assigned a height of 100 meters. Local agents (e.g., senior centers, neighborhood associations), with medium affluence, were assigned a height of 50 meters. Local stores (e.g., supermarkets, pharmacies, small businesses), categorized by affluence level: high affluence stores were assigned at 30 meters and low affluence stores at 10 meters. Figure \ref{fig:uoa1}c presents a topographic map with iso-lines representing varying altitudes based on affluence. The neighborhood's topography is predominantly low, with steep slopes in the north and southeast where socio-cultural and local agents infrastructures are located.

\subsection*{Pedestrian mobility definitions}\label{ss:definitions}
To characterize the pedestrian mobility of each group, we first define $\vec{r}(t)$ as the GPS two-coordinate vector of a pedestrian position at time $t$. Time duration of the lapse between consecutive GPS timestamps at time $t$ is $\Delta_t$ and it is irregular. The total distance $D$ and duration $T$ and of one trajectory thus read
\begin{equation}
D=\sum_{\{t\}} |\vec{r}(t+\Delta_t)-\vec{r}(t)|,
\label{eq:D}
\end{equation}
and 
\begin{equation}
T=\sum_{\{t\}} \Delta_t,
\label{eq:T}
\end{equation}
where we sum over all collection of timestamps $t$ of a trajectory (last timestep excluded). We can now discriminate two different states in the trajectories. Following the discussion in the Data Descriptor \cite{Larroya_2024}, first state is when pedestrians are moving and it will only consider those lapses smaller than 10 seconds ( $\Delta^m_t< 10 s$). Therefore we will take the range of values constrained between $\Delta = 1$s and $\Delta = 9$s. Under this condition, we can obtain the average time lapse between two consecutive GPS records: $E[\Delta_t^m] = 5.09 \pm 0.04$s. We can also define instantaneous velocity as
\begin{equation}
v(t^m)=\frac{|\vec{r}(t^m+\Delta^m_t)-\vec{r}(t^m)|}{\Delta^m_t}.
\label{eq:v}
\end{equation}
We can alternatively consider the state when pedestrian is stopped at if $\Delta (t) \geq 10$s, the GPS record at time $t$ corresponds to a stop. In that case, we can compute the stop durations as the timestamp difference between first GPS record labeled as ``stop'' and the last consecutive GPS record labeled as ``stop''. The statistical values are reported in Table \ref{tab:basic_statistics} and for the time duration we will use variable $\tau$.

\subsection*{Weighted $K$-means clustering algorithm}
 We will use clustering to group the action stops in different areas of the neighborhood. $K$-means algorithm works with the Euclidean distance between the points to be clustered. The algorithm chooses $K$ centroids (being $K$ the number of desired clusters) and assigns each point to a cluster depending on the distance to the different centroids. It then recalculates the position of the centroids until they converge, when the positions of the centroids do no longer change. The $K$-means++ initialization technique \cite{kmeans++} is used to choose the initial $K$ centroids at the initialization of the algorithm. The method selects the initial centroids using sampling based on an empirical probability distribution of the contribution of the records to the overall inertia. In this way, the initial selection is not completely random, and correct results are guaranteed while accelerating convergence. The algorithm also performs several iterations with different initial centroids, so that the final result does not depend on the initial conditions. We here perform $100$ iterations of the algorithm. In the case of action stops, the geolocated points are represented by its GPS coordinates (latitude and longitude). Therefore, we first convert the GPS coordinates (in degrees) to metres (x-y plane) to operate with Euclidean distances. For this, we use a projection to the corresponding UTM zone \cite{utm_proj}. We here use the weighted version for the $K$-means algorithm, which takes into account the time duration of each pausing event. In the algorithm, the duration (in seconds) represents the frequency of each data sample. Therefore, when recalculating the position of the centroids, longer stops have a greater impact. Figure \ref{fig:unweighted_k7} shows the resulting clusters for the unweighted $K$-means algorithm on the neighborhood map and they are compared with weighted $K$-means algorithm. Clusters are essentially the same, only one of the stops is regrouped in another cluster and the statistical features for each cluster remains very similar as shown in Table \ref{tab:clusters_unweighted_kmeans}. The duration of the stops when grouping them into clusters with the $K$-means algorithm does not seem to be a crucial element in the analysis. It is also possible to explore other clustering algorithms. The density-based DBSCAN algorithm \cite{Ester_1996} is applied to the data (see Figure \ref{fig:dbscan} and Table \ref{tab:dbscan_parameters}), giving worse results, being ineffective able to distinguish different high-density areas as $K$-means does. One might also question whether Euclidean distance is adequate for grouping stops based on proximity, given the presence of buildings and other urban elements. However, we found that typical routing services tend to overestimate path lengths compared to our walking routes. Pedestrians often take advantage of intermediate and pedestrian walkways characteristic of this neighborhood—features that routing services may not detect. For this reason, and given the small scale of the distances we work with, we consider Euclidean distance a suitable and sufficient option for classifying stops into high-density groups.

\subsection*{Number of optimal clusters}
$K$-means algorithm requires prefixing the number of clusters into which the data will be grouped. This raises the question of what is the optimal number of $K$ clusters. Here we apply three common techniques that can help determine the optimal number of clusters. 

\subsubsection*{Elbow method}
The Elbow method \cite{Bholowalia_2014} consists in computing, for each iteration of the weighted $K$-means algorithm with a number of clusters $K$, the within-cluster sum of squares $WCSS$, defined as 
\begin{equation}
    WCSS(K) = \sum_{k=1}^{K} \sum_{x_{i}\in C_{k}} \vert x_{i} - \mu_{k} \vert^{2},
    \label{eq:wcss}
\end{equation}
where $C_{k}$ represents the set of samples in cluster $k$, $x_{i}$ is each sample (position) of cluster $C_{k}$ and $\mu_{k}$ is the position of the centroid of cluster $C_{k}$. $WCSS$ thus measures the sum of the squared distances between samples in the clusters and their cluster centroids. Then, $WCSS(K)$ is represented as function of the number of clusters chosen $K$. The result is a decreasing curve, since the more clusters there are, the shorter the within-cluster distances. The optimal number of clusters is usually chosen as the value of $K$ from in which an elbow is observed, and the $WCSS$ stops decreases sharply. It is clear that this technique is rather visual and somewhat subjective, and may not lead to a clear or unique choice of a value of $K$. Figure \ref{fig:elbow} displays $WCSS$ as a function of the number of clusters, from $K=1$ to $K=9$, applying the weighted $K$-means algorithm to the stops data. We observe that there is no clear elbow point, at which the $WCSS$ value stops decreasing abruptly and stabilizes. It seems, however, that a value of $K$ between 5 and 7 could be a reasonable choice, since further increasing the number of clusters does not lead to a reduction of the $WCSS$. However, a more accurate and objective method needs to be applied to determine the optimal number of clusters and that is why we implement the Silhouette method.

\subsubsection*{Silhouette method}
The Silhouette analysis \cite{Rousseeuw_1987} evaluates the separation between clusters. It measures how close each point in a cluster is to points in neighboring clusters. It is defined as
\begin{equation}
\begin{cases}
    s_{i}= \frac{b_{i}-a_{i}}{\max\left(a_{i},b_{i}\right)} & \text{if} \quad N_{k} > 1 \\
    s_{i} = 0 & \text{if} \quad N_{k} = 1,  
    \end{cases}
    \label{si}
\end{equation}
where $N_{k}$ is the number of samples in cluster $C_{k}$. $a_{i}$ is, for each sample $x_{i}$ in cluster $C_{k}$, the mean distance between sample $x_{i}$ and all other samples in the same cluster $x_{j}$. It reads as 
\begin{equation}
    a_{i}= \frac{1}{N_{k}-1}\sum_{j\in C_{k}, i\neq j}\vert x_{i}-x_{j}  \vert.
    \label{ai}
\end{equation}
Finally, $b_{i}$ is the smallest mean distance of sample $x_{i}$ to all samples in any other cluster of which $x_{i}$ is not a member. It can be written as
\begin{equation}
    b_{i}= \min_{k\neq q}\frac{1}{N_{q}}\sum_{j\in C_{q}} \vert  x_{i}-x_{j}  \vert,
    \label{bi}
\end{equation}
where $N_{q}$ is the number of samples of cluster $C_{q}$ and $x_{j}$ are the samples of cluster $C_{q}$, which is the one with the smallest mean distance to sample $x_{i}$ from cluster $C_{k}$.

The range of $s_{i}$ is $\left[-1, 1\right]$. Silhouette coefficients near $1$ indicate that the sample is far away from the neighboring clusters. A value of 0 indicates that the sample is on or very close to the decision boundary between two neighboring clusters and negative values indicate that those samples might have been assigned to the wrong cluster. The optimal number of clusters $K$, is the one for which the average Silhouette value (over all data samples), $\langle s_{i}\rangle$, is maximised. In addition, it is convenient to visually analyse the silhouette coefficients $s_{i}$ of each cluster, to observe and verify that all clusters have broadly similar coefficients (for example, there are not too many negative values or some clusters with most of their coefficients below the mean value $\langle s_{i}\rangle $). Figure \ref{fig:average_silhouette} displays the average Silhouette scores for each number of chosen clusters $K$, from $K=2$ to $K=10$. The best score obtained is $\langle s_{i}\rangle = 0.52$ with $K=7$. Figure \ref{fig:silhouette} shows the results for each number of clusters chosen, together with the graphical representation of the clusters found. It confirms that indeed $K=7$ seems to be a reasonable choose for the number of optimal clusters.

\subsubsection*{GAP statistic method}
The last technique applied for robustness is the the GAP statistic \cite{Tibshirani_2001}. It compares the total within-cluster dispersion for different values of the number of clusters $K$ with their expected values under null reference distribution of the data (random uniform distribution). The $GAP(K)$ value is averaged over $B$ reference random datasets. It is defined as 
\begin{equation}
    GAP(K)= \frac{1}{B}\sum_{b=1}^{B} \log{(W_{bK})}-\log{(W_{K})}, 
    \label{gap}
\end{equation}
where $W_{K}=WCSS(K)$ is the within-cluster sum of squares defined above in Eq. (\ref{eq:wcss}) and $W_{bK}$ the same for the created data pool $b$. $B$ is the total number of $b$ reference generated data sets. The optimal number of clusters is the smallest $K$ such the $GAP(K)$ is maximized, i.e., is within one standard deviation of the $GAP(K+1)$. This means that the clustering structure is far away from the random uniform distribution of points. That is
\begin{equation}
    K: \quad GAP(K)\geq GAP(K+1)-\sigma_{K+1}.
    \label{gap_k}
\end{equation}
Figure \ref{fig:GAP} displays the value of $GAP(K)$ as a function of the number of clusters $K$, for $K=1$ to $K=9$ and using $B=500$ reference data sets. We find that the optimal number of clusters is $K=7$, which is also consistent with the Silhouette method. We also performed 50 iterations of the $GAP(K)$ calculation for robustness, giving always a result of $K=7$ as the optimal number of clusters.

\section*{Acknowledgements}

We thank the 72 volunteers from the school “Escola Ferrer i Guàrdia (EFiG)”, the associations “Associació de Veïns Sota el Cam\'i Ral” and “Espai Actiu de la Gent Gran” for their participation. The study was partially supported by the Ministerio de Ciencia e Innovación and Agencia Estatal de Investigación MCIN/AEI/10.13039/501100011033, grant number PID2019-106811GB-C33 (FL and JP); by MCIN/AEI/10.13039/501100011033 and by “ERDF A way of making Europe", grant number PID2022-140757NB-I00 (FL and JP). We also acknowledge the support of Generalitat de Catalunya through Complexity Lab Barcelona, grant number 2021SGR00856 (FL, JP); and by the ICI programme (Interculturalitat i Cohesió Social) funded by Fundació Bancària “la Caixa”.

\section*{Author contributions statement}

All authors conceived the experiment, all authors conducted the experiment(s), F.L. and J.P. analysed the results, all authors discussed the results.  All authors reviewed the manuscript. 

\section*{Additional information}

\textbf{Accession codes}. The analysis was performed using the Python programming language (in particular, the Web-based interactive development environment Jupyter Notebook). The code to process the input data and reproduce our main results and figures is publicly available on GitHub \url{https://github.com/ferranlarroyaub/Pedestrian_mobility_Granollers}.

\noindent \textbf{Competing interests}. The authors declare no competing interests.

\noindent \textbf{Ethics declaration} The Universitat de Barcelona Ethics Committee (IRB00003099) has approved this mobility experiment. All participants read and signed the informed consent. As we did not collect any personal data, no privacy issues are in conflict with full public release of the underlying data.

\AddToHook{enddocument/afteraux}{%
\immediate\write18{
cp output.aux MAIN.aux
}%
}

\begin{thebibliography}{1}
\expandafter\ifx\csname url\endcsname\relax
  \def\url#1{\texttt{#1}}\fi
\expandafter\ifx\csname urlprefix\endcsname\relax\def\urlprefix{URL }\fi
\providecommand{\bibinfo}[2]{#2}
\providecommand{\eprint}[2][]{\url{#2}}

\bibitem{Batty_2013} Batty, M. {\it The New Science of Cities} (MIT Press, 2013). \url{https://doi.org/10.7551/mitpress/9399.001.0001}

\bibitem{Barthelemy_2016} Barthelemy, M. {\it The Structure and Dynamics of Cities} (Cambridge Univ. Press, 2016). \url{https://doi.org/10.1017/9781316271377}

\bibitem{Bettencourt_2021} Bettencourt, L.M.A. {\it Introduction to Urban Science: Evidence and Theory of Cities as Complex Systems} (MIT Press, 2021). \url{https://doi.org/10.7551/mitpress/13909.001.0001}

\bibitem{Gravier_2024} Gravier, J. \& Barthelemy, M. A typology of activities over a century of urban growth. {\it Nat Cities} 1, 567–575 (2024). \url{https://doi.org/10.1038/s44284-024-00108-7}

\bibitem{Verbavatz_2020} Verbavatz, V. \& Barthelemy, M. The growth equation of cities. {\it Nature} 587, 397-–401 (2020). \url{https://doi.org/10.1038/s41586-020-2900-x}

\bibitem{Henderson_2007} Henderson, J.V. \& Wang, H.-G. Urbanization and city growth: the role of institutions. {\it Reg Sci Urban Econ} 37, 283–313 (2007). \url{https://doi.org/10.1016/j.regsciurbeco.2006.11.008}

\bibitem{Barthelemy_2013} Barthelemy, M., Bordin, P., Berestycki, H. et al. Self-organization versus top-down planning in the evolution of a city. {\it Sci Rep} 3, 2153 (2013). \url{https://doi.org/10.1038/srep02153}

\bibitem{Bettencourt_2007} Bettencourt, L.M.A., Lobo, J., Helbing, D., Kühnert, C., \& West, G.B. Growth, innovation, scaling, and the pace of life in cities. {\it Proc Natl Acad Sci USA} 
104 (17), 7301-7306 (2007).
\url{https://doi.org/10.1073/pnas.0610172104}

\bibitem{Sennett_2020} Sennett, R. \& Sendra, P. {\it Designing disorder: Experiments and disruptions in the city} (Verso Books, 2020).


\bibitem{UN_Habitat_2023} Ignatova, A. {\it My Neighboorhood} (UN-Habitat, 2023). Report
\url{https://unhabitat.org/sites/default/files/2023/05/my_neighbourhood_publication_19.05.2359.pdf}

\bibitem{Jackson_2003} Jackson, L. E. The relationship of urban design to human health and condition. {\it Landsc Urban Plan} 64(4), 191-200 (2003). \url{https://doi.org/10.1016/S0169-2046(02)00230-X}

\bibitem{Mazumdar_2018} Mazumdar, S., Learnihan, V., Cochrane, T., \& Davey, R. The Built Environment and Social Capital: A Systematic Review. {\it Environ Behav} 50(2), 119-158 (2018). \url{https://doi.org/10.1177/0013916516687343}

\bibitem{Carmona_2019} Carmona, M. Principles for public space design, planning to do better. {\it Urban Des Int} 24, 47–59 (2019). \url{https://doi.org/10.1057/s41289-018-0070-3}

\bibitem{Sonta_2023} Sonta, A., \& Jiang, X. Rethinking walkability: Exploring the relationship between urban form and neighborhood social cohesion. {\it Sustain Cities Soc} 99, 104903 (2023). \url{https://doi.org/10.1016/j.scs.2023.104903}

\bibitem{Parker_2023} Parker, G., Wargent, M., Salter, K. \& Yuille, A. Neighbourhood planning in England: A decade of institutional learning. {\it 
Progress in Planning} 174, 100749 (2023). \url{https://doi.org/10.1016/j.progress.2023.100749}

\bibitem{Healey_1998} Healey, P. Collaborative planning in a stakeholder society. {\it Town Planning Review} 69 (1), 1--21 (1998). \url{https://doi.org/10.3828/tpr.69.1.h651u2327m86326p}

\bibitem{Gehl_2013} Gehl, J. {\it Cities for people} (Island press, 2013).

\bibitem{Rodela_2025} Rodela, R., Williams, M., Ohlsson, J. et al. Six propositions for care-centric planning and governance that promote sustainable cities. {\it npj Urban Sustain} 5, 21 (2025). \url{https://doi.org/10.1038/s42949-025-00214-y}

\bibitem{Szaboova_2024} Szaboova, L. Adger, W.N., Safra de Campos, R. et al. Promoting sustainable cities through creating social empathy between new urban populations and planners. {\it npj Urban Sustain} 4, 52 (2024). \url{https://doi.org/10.1038/s42949-024-00189-2}

\bibitem{pps_web} Project for Public Spaces. (2007). What Is Placemaking? 
Center for Transformative Placemaking. New York, NY. \url{https://www.pps.org/article/what-is-placemaking} Accessed 23 September 2024.

\bibitem{EPA_2021} EPA. 2021. Climate Change and Social Vulnerability in the United States: A Focus on Six Impacts. U.S. Environmental Protection Agency, EPA 430-R-21-003. \url{https://www.epa.gov/cira/social-vulnerability-report}

\bibitem{EU_2020} Fioretti, C., Pertoldi, M., Busti, M. \& Van Heerden, S. (eds), Handbook of Sustainable Urban Development Strategies, EUR 29990 EN, Publications Office of the European Union, Luxembourg, 2020, ISBN 978-92-76-16425-8, \url{https://doi.org/10.2760/020656}, JRC118841.

\bibitem{Thomas_2016} Thomas, D. Placemaking: An urban design methodology (Routledge, New York, 2016). \url{https://doi.org/10.4324/9781315648125}

\bibitem{Ellery_2021} Ellery, P.J., Ellery, J. \& Borkowsky, M. Toward a Theoretical Understanding of Placemaking. {\it Int Journal of Com WB} 4, 55–76 (2021). \url{https://doi.org/10.1007/s42413-020-00078-3}

\bibitem{Palmer_2024} Palmer, L. New inroads on community-centric placemaking. {\it Nat Cities} 1, 2–4 (2024). \url{https://doi.org/10.1038/s44284-023-00015-3}

\bibitem{Sandercock_2024} Sandercock, L. Reimagining the soul of urban planning. {\it Nat Cities} 1, 5–6 (2024). \url{https://doi.org/10.1038/s44284-023-00010-8}

\bibitem{Toolis_2017} Toolis, E. E. Theorizing critical placemaking as a tool for reclaiming public space. {\it Am J Community Psychol} 59, 184--199 (2017). \url{https://doi.org/10.1002/ajcp.12118}

\bibitem{Omholt_2019} Omholt, T. Strategies for inclusive place making. {\it J. Place Manag. Dev.} 12(1), 2-19 (2019). \url{https://doi.org/10.1108/JPMD-09-2017-0098}

\bibitem{Couper_2023} Couper, I., Jaques, K., Reid, A., \& Harris, P. Placemaking and infrastructure through the lens of levelling up for health equity: A scoping review. {\it Health \& Place} 80, 102975 (2023). \url{http://doi.org/10.1016/j.healthplace.2023.102975}

\bibitem{Lorono_2023} Loroño-Leturiondo, M. \& Illingworth, S. Gender and placemaking: talking to women about clean air and sustainable urban environments in changing cities. {\it J Place Manag Dev} 16(1), 91--104 (2023). \url{https://doi.org/10.1108/jpmd-04-2021-0035}

\bibitem{Sutton_2002} Sutton, S.E., \& Kemp, S.P. Children as partners in neighborhood placemaking: Lessons from intergenerational design charrettes. {\it J Environ Psychol} 22(1-2), 171--189 (2002). \url{https://doi.org/10.1006/jevp.2001.0251}

\bibitem{Lager_2021} Lager, D. R., Van Hoven, B. \& Huigen, P. P. P.  Neighbourhood walks as place-making in later life. {\it  Soc Cult Geogr} 22(8), 1080--1098 (2021). \url{https://doi.org/10.1080/14649365.2019.1672777}

\bibitem{Yildiz_2024} Yıldız, A., Gürel, A., Gürel, M.Ö.
Migrants as agents in placemaking: A socio-spatial analysis of Basmane area in Izmir, Türkiye, {\it Wellbeing Space Soc} 7, 100222 (2024). \url{https://doi.org/10.1016/j.wss.2024.100222}

\bibitem{Gulsrud_2018} Gulsrud, N.M., Hertzog, K., \& Shears, I. Innovative urban forestry governance in Melbourne?: Investigating “green placemaking” as a nature-based solution. {\it Environ Res} 161, 158--167 (2018). \url{https://doi.org/10.1016/j.envres.2017.11.005}

\bibitem{Boros_2021} Boros, J., \& Mahmoud, I. Urban design and the role of placemaking in mainstreaming nature-based solutions. Learning from the Biblioteca Degli Alberi case study in Milan. {\it Front Sustain Cities} 3, 635610 (2021). \url{https://doi.org/10.3389/frsc.2021.635610}

\bibitem{Ferster_2017} Ferster, C., Nelson, T., Laberee, K., Vanlaar, W. \& Winters, M. Promoting crowdsourcing for urban research: Cycling safety citizen science in four cities. {\it Urban Sci} 1(2), 21 (2017). \url{https://doi.org/10.3390/urbansci1020021}


\bibitem{Acuto_2018} Acuto, M., Parnell, S., \& Seto, K. C. Building a global urban science. {\it Nat Sustain} 1 2-4 (2018).  \url{https://doi.org/10.1038/s41893-017-0013-9}

\bibitem{Harley_2002} Harley, J.B. {\it The new nature of maps: essays in the history of cartography} (JHU Press, 2002). \url{https://doi.org/10.1093/ehr/118.478.1093}

\bibitem{Sandercock_2023} Sandercock, L. {\it Mapping possibility: Finding purpose and hope in community planning} (Routledge, London, 2023).

\bibitem{Paez_2024a} Paez, R. {\it Operative Mapping: The Use of Maps as a Design Tool} (Actar D, 2024).

\bibitem{Paez_2024b} Paez, R., Valtchanova, M., Larroya, F. \& Perelló, J. Maps as design tools: Space, time and experience. In: Rossetto, T. \& Lo Presti, L. The Routledge Handbook of Cartographic Humanities, pp. 172-181 (Routledge, 2024). \url{https://doi.org/10.4324/9781003327578}

\bibitem{Xu_2023} Xu, Y., Olmos, L.E., Mateo, D. et al. Urban dynamics through the lens of human mobility. {\it Nat Comput Sci} 3, 611–620 (2023).
\url{https://doi.org/10.1038/s43588-023-00484-5}

\bibitem{Moreno_2021} Moreno C., Allam Z., Chabaud D., Gall C., Pratlong, F. Introducing the “15-Minute City”: Sustainability, Resilience and Place Identity in Future Post-Pandemic Cities. {\it Smart Cities} 4(1), 93--111 (2021). \url{https://doi.org/10.3390/smartcities4010006}

\bibitem{Bruno_2024} Bruno, M., Monteiro Melo, H.P., Campanelli, B. et al. A universal framework for inclusive 15-minute cities. {\it Nat Cities} 1, 633–641 (2024). \url{https://doi.org/10.1038/s44284-024-00119-4}

\bibitem{Abbiasov_2024} Abbiasov, T., Heine, C., Sabouri, S. et al. The 15-minute city quantified using human mobility data. {\it Nat Hum Behav} 8, 445–455 (2024). \url{https://doi.org/10.1038/s41562-023-01770-y}


\bibitem{Arvidsson_2023} Arvidsson, M., Lovsjö, N. \& Keuschnigg, M. Urban scaling laws arise from within-city inequalities. {\it Nat Hum Behav} 7, 365-–374 (2023). \url{https://doi.org/10.1038/s41562-022-01509-1}

\bibitem{Xu_2025} Xu, F., Wang, Q., Moro, E. et al. Using human mobility data to quantify experienced urban inequalities. {\it Nat Hum Behav} 9, 654–664 (2025). \url{https://doi.org/10.1038/s41562-024-02079-0}

\bibitem{Rhoads_2023b} Rhoads, D., Rames, C., Solé-Ribalta, A., González, M. C., Szell, M., \& Borge-Holthoefer, J. Sidewalk networks: Review and outlook.  {\it Comput Environ Urban Syst} 106, 102031 (2023). \url{https://doi.org/10.1016/j.compenvurbsys.2023.102031}

\bibitem{Helbing_2001} Helbing, D., Molnár, P., Farkas, I. J., \& Bolay, K. Self-Organizing Pedestrian Movement. {\it Environ. Plann. B Plann. Des.} 28(3), 361--383 (2001). \url{https://doi.org/10.1068/b2697}

\bibitem{Jiang_2016} Jiang, S. et al. The TimeGeo modeling framework for urban mobility without travel surveys. {\it Proc Natl Acad Sci USA} 113, E5370–E5378 (2016). \url{https://doi.org/10.1073/pnas.1524261113}

\bibitem{Bongiorno_2021} Bongiorno, C., Zhou, Y., Kryven, M. et al. Vector-based pedestrian navigation in cities. {\it Nat Comput Sci} 1, 678–685 (2021). \url{https://doi.org/10.1038/s43588-021-00130-y}

\bibitem{Hunter_2021} Hunter, R.F., Garcia, L., de Sa, T.H. et al. Effect of COVID-19 response policies on walking behavior in US cities. {\it Nat Commun} 12, 3652 (2021). \url{https://doi.org/10.1038/s41467-021-23937-9}

\bibitem{Yabe_2024} Yabe, T., Luca, M., Tsubouchi, K. et al. Enhancing human mobility research with open and standardized datasets. {\it Nat Comput Sci} 4, 469–472 (2024). \url{https://doi.org/10.1038/s43588-024-00650-3}

\bibitem{Yabe_2024b} Yabe, T., Tsubouchi, K., Shimizu, T. et al. YJMob100K: City-scale and longitudinal dataset of anonymized human mobility trajectories. {\it Sci Data} 11, 397 (2024). \url{https://doi.org/10.1038/s41597-024-03237-9}

\bibitem{Perello_2024} Perelló, J., Larroya, F., Bonhoure, I., Peter, F. Citizen science for social physics: digital tools and participation. {\it Eur Phys J Plus} 139(7), 572 (2024). \url{https://doi.org/10.1140/epjp/s13360-024-05336-3}

\bibitem{Larroya_2023a} Larroya, F., Díaz, O., Sagarra, O. et al. Home-to-school pedestrian mobility GPS data from a citizen science experiment in the Barcelona area. {\it Sci Data} 10, 428 (2023). \url{https://doi.org/10.1038/s41597-023-02328-3}


\bibitem{Haklay_2012} Haklay, M. in Crowdsourcing geographic knowledge (eds. Sui, D., Elwood, S. \& Goodchild, M.) Chapter: Citizen science and volunteered geographic information: Overview and typology of participation pp. 105–122 (Springer, 2012).

\bibitem{Vohland_2021} Vohland, K., Land-Zandstra, A., Ceccaroni, L., Lemmens, R., Perelló, J., Ponti, M., Samson, R., \& Wagenknecht, K. {\it The Science of Citizen Science} (Springer, Cham, 2021).

\bibitem{Haklay_2021} Haklay, M. etal. Contours of citizen science: a vignette study, {\it Royal Society Open Science} 8, 202108 (2021)
\url{https://doi.org/10.1098/rsos.202108}

\bibitem{Cooper_2021} Cooper, C.B. et al. Inclusion in citizen science: The conundrum of rebranding. {\it Science} 372, 1386--1388 (2021). \url{ https://doi.org/10.1126/science.abi6487}

\bibitem{Bonhoure_2024} Bonhoure, I., Guba, B., Peer, C., Labastida, I., \& Perelló, J. (2025). Citizen science contributions to sustainable urban transformation and urban sustainability: a systematic literature review. {\it SocArXiv} zc7w4. \url{https://doi.org/10.31235/osf.io/zc7w4_v1}

\bibitem{Sagarra_2016} Sagarra, O., Gutiérrez-Roig, M., Bonhoure, I., \& Perelló J. Citizen Science Practices for Computational Social Science Research: The Conceptualization of Pop-Up Experiments. {\it Front Phys} 3, 93 (2016).  \url{https://doi.org/10.3389/fphy.2015.00093}

\bibitem{Perello_2022} Perelló, J. New knowledge environments: On the possibility of a citizen social science. {\it Metode Science Studies Journal} 12, 25-31 (2022). \url{https://doi.org/10.7203/metode.12.18136}

\bibitem{Barbosa_2018} Barbosa, H. et al. Human mobility: Models and applications. {\it Phys Rep} 734, 1–74 (2018). \url{https://doi.org/10.1016/j.physrep.2018.01.001}

\bibitem{Janez_2022} Janež, M, Verovšek, Š., Zupančič, T. \& Moškon, M. Citizen science for traffic monitoring: investigating the potentials for complementing traffic counters with crowdsourced data. {\it Sustainability} 14(2), 2 (2022) \url{https://doi.org/10.3390/su14020622}

\bibitem{Keseru_2019} Keserü, I., Wuytens, N., Macharis, C. Citizen observatory for mobility: a conceptual framework. {\it Transp Rev} 39(4), 485–510 (2019). \url{https://doi.org/10.1080/01441647.2018.1536089}

\bibitem{Pappers_2022} Pappers, J., Keserü, I., De Wilde, L., Evaluating citizen science data: a citizen observatory to measure cyclists’ waiting times. {\it Transp Res Interdiscip Perspect} 14, 100624 (2022). \url{https://doi.org/10.1016/j.trip.2022.100624}

\bibitem{Gutierrez_2016} Gutiérrez-Roig, M., Sagarra, O., Oltra, A., Palmer, J.R.B., Bartumeus, F., Diaz-Guilera, A. \& Perelló, J. Active and reactive behaviour in human mobility: the influence of attraction points on pedestrians. {\it Royal Society Open Science} 3, 160177 (2016). \url{https://doi.org/10.1098/rsos.160177}

\bibitem{Kapenekakis_2017} Kapenekakis, I. \& Chorianopoulos, K. Citizen science for pedestrian cartography: collection and moderation of walkable routes in cities through mobile gamification. {\it Hum. Cent. Comput. Inf. Sci.} 7, 10 (2017). \url{https://doi.org/10.1186/s13673-017-0090-9}

\bibitem{Ertz_2021} Ertz, O., Fischer, A. Ghorbel, H., Hüsser, O., Sandoz, R. \& Scius-Bertrand, A. Citizen participation \& digital tools to improve pedestrian mobility in cities. {\it Int. Arch. Photogramm. Remote Sens. Spatial Inf. Sci.} 46, 29–34 (2021). \url{https://doi.org/10.5194/isprs-archives-XLVI-4-W1-2021-29-2021}

\bibitem{Mueller_2018} Mueller, J., Lu, H., Chirkin, A., Klein, B. \& Schmitt, G. Citizen Design Science: A strategy for crowd-creative urban design. {\it Cities} 72, 181--188 (2018).
\url{https://doi.org/10.1016/j.cities.2017.08.018}

\bibitem{Christine_2021} Christine, D.I., \& Thinyane, M. Citizen science as a data-based practice: A consideration of data justice. {\it Patterns} 2(4) (2021). \url{https://doi.org/10.1016/j.patter.2021.100224}


\bibitem{Toomey_2020} Toomey, A.H., Strehlau-Howay,  L., Manzolillo, B. \& Thomas, C. 
The place-making potential of citizen science: Creating social-ecological connections in an urbanized world, {\it Landsc. Urban Plan.} 200, 103824 (2020).
\url{https://doi.org/10.1016/j.landurbplan.2020.103824}

\bibitem{Ramirez_2014} Ramirez-Andreotta, M., Brusseau, M.L., Artiola, J.F., Maier, R.M. \&  Gandolfi, A.J. Environmental Research Translation: Enhancing interactions with communities at contaminated sites, {\it Sci. Total Environ.} 497–498, 651-664 (2014) \url{https://doi.org/10.1016/j.scitotenv.2014.08.021}

\bibitem{Croese_2021} Croese, S., Dominique, M. \& Raimundo, I.M. Co-producing urban knowledge in Angola and Mozambique: towards meeting SDG 11. {\it npj Urban Sustain} 1, 8 (2021). \url{https://doi.org/10.1038/s42949-020-00006-6}

\bibitem{Grootjans_2022} Grootjans, S.J.M., Stijnen, M.M.N., Kroese, M.E.A.L., Ruwaard, D. \& Jansen, I.M.W.J. Citizen science in the community: Gaining insight in community and participant health in four deprived neighbourhoods in the Netherlands,
{\it Health \& Place} 75, 102798 (2022).
\url{https://doi.org/10.1016/j.healthplace.2022.102798}

\bibitem{Albert_2021} Albert, A., Bal\'azs, B., Butkevičien˙ e, E., Mayer, K., \& Perell\'o, J. (2021). Citizen Social Science: New and Established Approaches to Participation in Social Research. In: Vohland, K., et al. The Science of Citizen Science Ch. 7 (Springer, Cham, 2021).
\url{https://doi.org/10.1007/978-3-030-58278-4_7}

\bibitem{Bonhoure_2023} Bonhoure, I., Cigarini, A., Vicens, J., Mitats, B., \& Perell\'o, J.. Reformulating computational social science with citizen social science: the case of a community-based mental health care research. {\it Humanit Soc Sci Commun} 10, 81 (2023).
\url{https://doi.org/10.1057/s41599-023-01577-2} 

\bibitem{Pitidis_2024} Pitidis, V., Coaffee, J. \& Lima-Silva, F. 
Advancing equitable ‘resilience imaginaries’ in the Global South through dialogical participatory mapping: Experiences from informal communities in Brazil, {\it Cities} 150, 105015 (2024). \url{https://doi.org/10.1016/j.cities.2024.105015}

\bibitem{Larroya_2024} Larroya, F., Paez, R., Valtchanova, M. \& Perell\'o. Explorative pedestrian mobility GPS data from a citizen science experiment in a neighbourhood. {\it arXiv preprint} arXiv:2410.08672 (2024). \url{https://doi.org/10.48550/arXiv.2410.08672}

\bibitem{Hartigan_1979} Hartigan, J. A. \& Wong, M. A. A k-means clustering algorithm. {\it J. R. Stat. Soc., C: Appl. Stat.} 28(1), 100-108 (1979). \url{https://doi.org/10.2307/2346830}

\bibitem{Bholowalia_2014} Bholowalia, P. \& Kumar, A. EBK-means: A clustering technique based on elbow method and k-means in WSN. {\it Int. J. Comput. Appl.} 105(9) (2014). \url{https://doi.org/10.5120/18405-9674}

\bibitem{Rousseeuw_1987} Rousseeuw, P.J. Silhouettes: a graphical aid to the interpretation and validation of cluster analysis. {\it J. Comput. Appl. Math.} 20, 53-65 (1987).
\url{https://doi.org/10.1016/0377-0427(87)90125-7}

\bibitem{Tibshirani_2001} Tibshirani, R., Walther, G., \& Hastie, T. Estimating the number of clusters in a data set via the gap statistic. {\it J. R. Stat. Soc., B: Stat. Methodol.} 63(2), 411-423 (2001). \url{https://doi.org/10.1111/1467-9868.00293}

\bibitem{Ravazzoli_2017} Ravazzoli, E., \& Torricelli, G. P. Urban mobility and public space. A challenge for the sustainable liveable city of the future. {\it J. Public Space} 2(2), 37--50 (2017).
\url{https://doi.org/10.5204/jps.v2i2.9}

\bibitem{Tobin_2022} Tobin, M., Hajna, S., Orychock, K. et al. Rethinking walkability and developing a conceptual definition of active living environments to guide research and practice. {\it BMC Public Health} 22, 450 (2022). \url{https://doi.org/10.1186/s12889-022-12747-3}

\bibitem{Paez_2024c} Paez, R. Design as Playground: Exploring Spatial Design Through Playful Practices. {\it Space and Culture} 27(2), 187--208 (2024). \url{https://doi.org/10.1177/12063312231213249}


\bibitem{Senabre_2018} Senabre, E., Ferran-Ferrer, N., \& Perelló, J. Participatory Design of Citizen Science Experiments. {\it Comunicar: Media Education Research Journal} 26(54), 29--38 (2018). \url{https://doi.org/10.3916/C54-2018-03}

\bibitem{Senabre_2021} Senabre Hidalgo, E., Perelló, J., Becker, F., Bonhoure, I., Legris, M. \& Cigarini, A. Participation and co-creation in citizen science." Chapter 11. In: Vohland K. et al.(Eds). 2021. The Science of Citizen Science. Springer pp: 199-218 (2021). \url{https://doi.org/10.1007/978-3-030-58278-4}

\bibitem{Wikiloc} \url{https://www.wikiloc.com} [Accessed: November 5, 2024]

\bibitem{Wang_2020} Wang, J., \& Kwan, M. P. Daily activity locations k-anonymity for the evaluation of disclosure risk of individual GPS datasets. {\it Int J Health Geogr} 19(1), 1-14 (2020).
\url{https://doi.org/10.1186/s12942-020-00201-9}

\bibitem{Larroya_2023b} Larroya, F. \& Perelló, J. Explorative pedestrian mobility GPS data from a citizen science experiment in a neighbourhood. CORA. Repositori de Dades de Recerca \url{https://doi.org/10.34810/data898} (2023).


\bibitem{kmeans++} Arthur, D., \& Vassilvitskii, S. k-means++: The advantages of careful seeding. {\it Stanford} (2006). \url{https://theory.stanford.edu/~sergei/papers/kMeansPP-soda.pdf}

\bibitem{utm_proj} Langley, R. B. The UTM grid system. {\it  GPS world} 9(2), 46-50 (1998). \url{http://131.202.94.44/papers.pdf/gpsworld.february98.pdf}

\bibitem{Ester_1996} Ester, M., Kriegel, H. P., Sander, J., \& Xu, X.  
A density-based algorithm for discovering clusters in large spatial databases with noise. {\it Proceedings of the Second International Conference on Knowledge Discovery and Data Mining (KDD'96)}, 226-231 (1996). \url{https://cdn.aaai.org/KDD/1996/KDD96-037.pdf}

\end{thebibliography}
\end{document}


\maketitle

\tableofcontents
\newpage
\listoffigures
\listoftables
\newpage

\section{Opportunity Units} \label{sec:uoa}
\begin{figure}[H]
\centering
\includegraphics[width=0.95\linewidth]{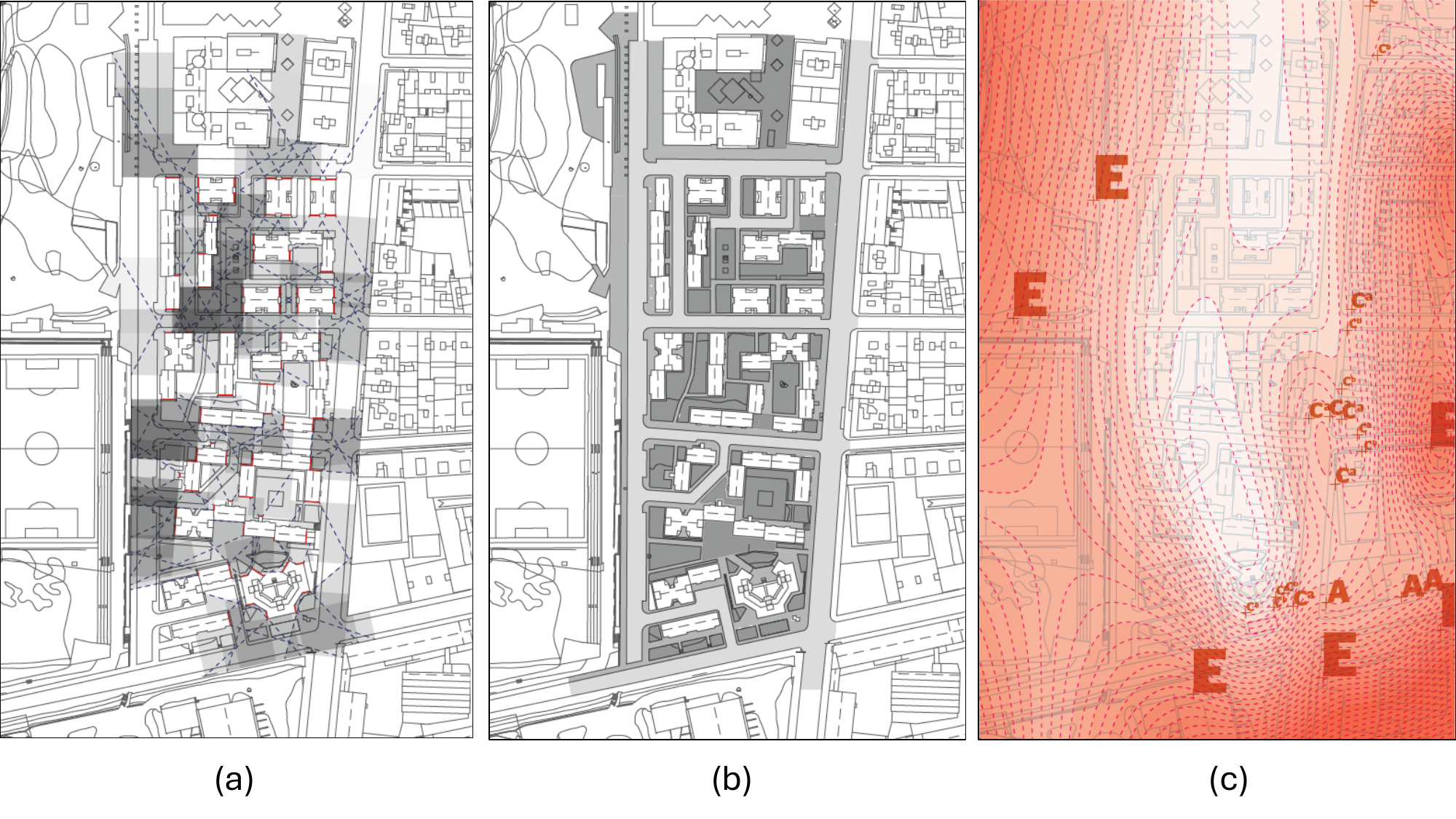}
\caption[Mapping procedure to determine the opportunity units.]{\textbf{Mapping procedure to determine the activation opportunity units.} (a) Visual dimension: Result of the interaction between urban objects in the neighborhood and the degree of visual vertical space generated by the blind facades of the buildings. The darker regions (where several visibility areas converge) are considered regions of potential interest. (b) Spatial dimension: There are four levels of appropiability depending on the degree of pedestrian accessibility of the public space: from zero appropiability (private space, in white) to high appropiability (intermediate spaces like squares and green flowerbeds, in 70\% black). (c) Social dimension: Topographic map with the attractors of socio-cultural activity in the neighborhood. The socio-cultural agents are considered attractors, and are assigned a certain height based on the degree of affluence of people they attract. There are three main groups: Education and cultural locations (high level of affluence), local agents (medium level of affluence) and local stores (high or low affluence).}
\label{fig:uoa1}
\end{figure}

Figure \ref{fig:uoa1} shows the mappings developed from a visual, spatial and socio-cultural approach that allowed the expert architects to obtain the eight opportunity units (OU) \cite{Paez_2024b,Valtchanova_2023}. Figure \ref{fig:uoa2} shows the result of the overlay of the three appropriability maps, resulting in the eight OU. Details of the procedure are described in the main paper.

\section{$K$-means clustering}\label{s:weighted_kmeans}

\subsection{Unweighted $K$-means}\label{ss:unweighted_kmeans}
\begin{figure}[H]
\centering
\includegraphics[width=0.99\linewidth]{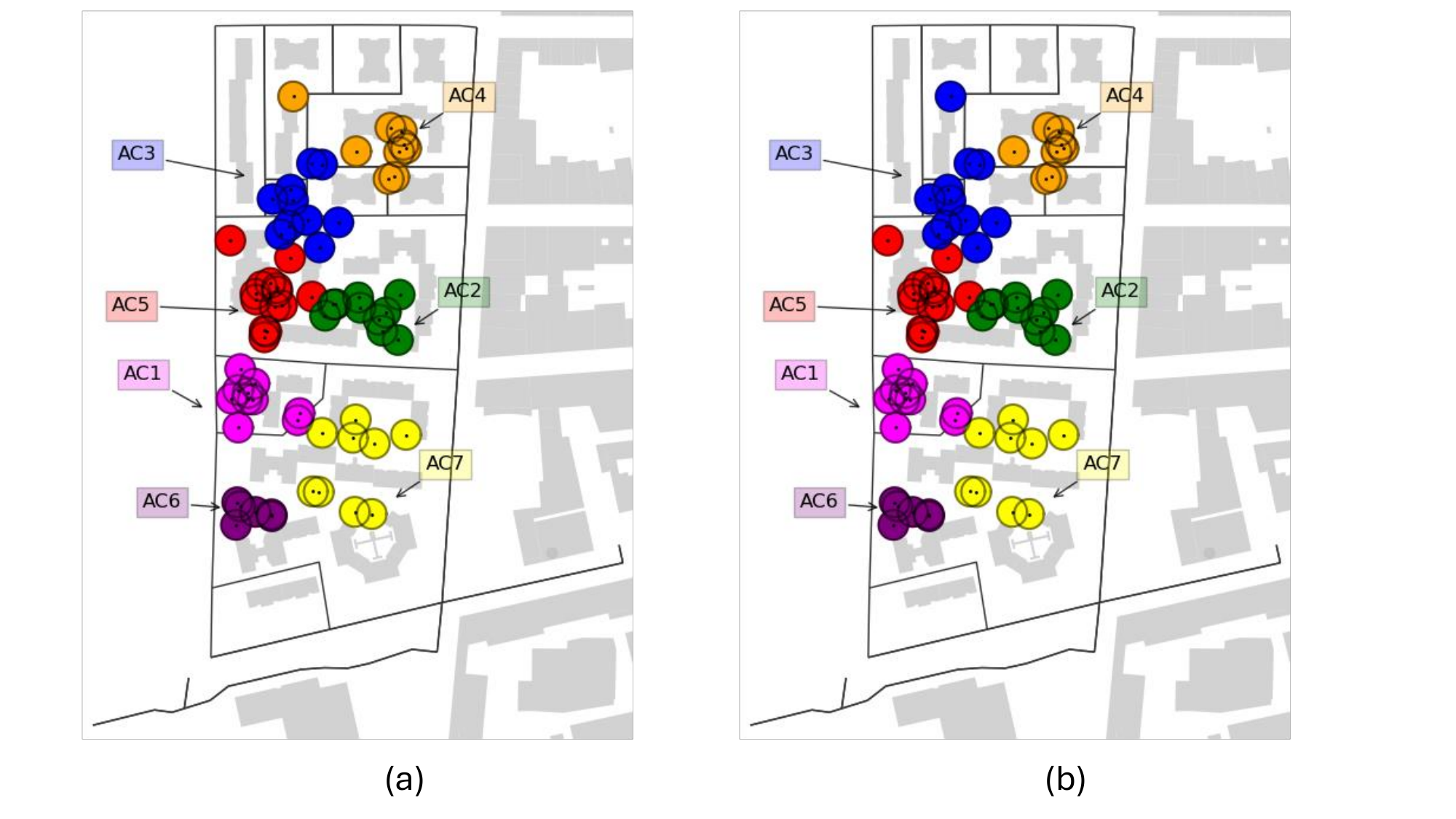}
\caption[Action stops within the 7 $K$-means clusters.]{\textbf{Action stops within the 7 $K$-means clusters.} The 7 appropriation clusters applying the $K$-means algorithm. (a) Weighted version, using the time duration of each stop as the weights. (b) Unweighted version. For visualization purposes, all stops are represented with bubbles of the same size, with each cluster of a different color.}
\label{fig:unweighted_k7}
\end{figure}

Using the classic $K$-means algorithm \cite{Hartigan_1979,kmeans++} (unweighted), we obtain essentially the same clusters as using the weighted version, in which each stop is weighted with its duration. Figure \ref{fig:unweighted_k7} compares the cluster representation in both cases. We observe that there is only one stop (the one located at the top of the map) that has changed from the AC3 activation cluster (in the weighted version, Figure \ref{fig:unweighted_k7}a) to AC4 (in the unweighted version, Figure \ref{fig:unweighted_k7}b). Table \ref{tab:clusters_unweighted_kmeans} displays the statistics of the stops of each activation cluster in the unweighted version. The statistics are essentially the same as in the weighted $K$-means (see Table \ref{tab:clusters_weighted_kmeans} in the main paper), and there are only minor differences in the AC3 and AC4 clusters which exchange one stop. 

\begin{table}[H]
\centering
\begin{adjustbox}{width=\textwidth}
\begin{tabular}{ccccccc}
\hline \hline
Cluster & Stops (\%) & Total $T_{stop}$ (s) & $\langle T_{stop} \rangle$ (s) & $\sigma_{T_{stop}}$ (s) & $T_{stop}^{min}$ (s) & $T_{stop}^{max}$ (s) \\ \hline
AC1 (magenta) & $10$ $(15\%)$ & $8,083$ $(15\%)$ & $808\pm 202$ & $638$ & $100$ & $1,955$\\
AC2 (green) & $10$ $(15\%)$ & $10,532$ $(20\%)$ & $1,053\pm 276$ & $872$ & $200$ & $3,130$ \\
AC3 (blue) & $11$ $(15\%)$ & $12,305$ $(22\%)$ & $1,118\pm 252$ & $834$ & $96$ & $2,677$ \\
AC4 (orange) & $8$ $(13\%)$ & $4,480$ $(9\%)$ & $560\pm 117$ & $332$ & $100$ & $963$ \\
AC5 (red) & $14$ $(20\%)$ & $7,761$ $(15\%)$ & $547\pm 150$ & $563$ & $87$ & $2,345$ \\
AC6 (purple) & $6$ $(9\%)$ & $4,349$ $(8\%)$ & $725\pm 162$ & $397$ & $262$ & $1,243$ \\
AC7 (yellow) & $9$ $(13\%)$ & $5,650$ $(11\%)$ & $628\pm 262$ & $785$ & $39$ & $2,669$ \\ \hline
All stops & $68$ $(100\%)$ & $53,060$ $(100\%)$ & $780\pm 83$ & $680$ & $39$ & $3,130$ \\
\hline\hline
\end{tabular}
\end{adjustbox}
\caption[Statistics of the 7 $K$-means unweighted activation clusters.]{\textbf{Statistics of the 7 $K$-means unweighted activation clusters.} Number of stops, total duration of stops, average stop duration, standard deviation and the shortest and the longest stop for each one of the 7 $K$-means activation clusters (unweighted case).}
\label{tab:clusters_unweighted_kmeans}
\end{table}

\subsection{Optimal $K$ number of clusters: Elbow, Silhouette and GAP statistic methods}\label{ss:optimalK}
\begin{figure}[H]
\centering
\includegraphics[width=0.80\linewidth]{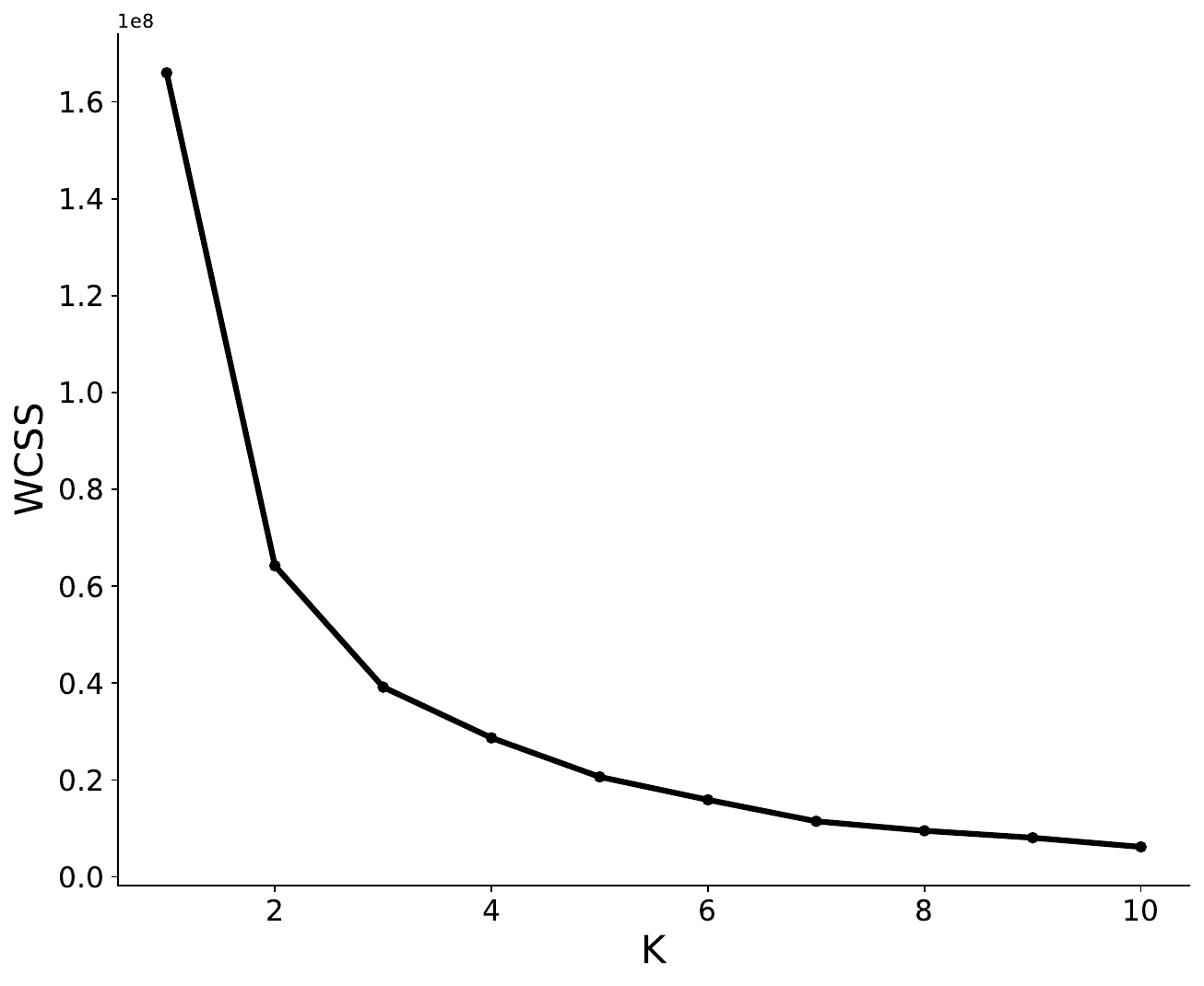}
\caption[Within-cluster sum of squares as a function of the number of clusters.]{\textbf{\bf Within-cluster sum of squares as a function of the number of clusters.} The shape of the within-cluster sum of squares (WCSS) curve shows that a value of $K$ between 5 and 7 could be a reasonable choice for the number of optimal clusters.}
\label{fig:elbow}
\end{figure}

The first metric we compute to study quality of the resulting clusters and therefore the number of optimal clusters is the within-cluster sum of squares (WCSS, Eq. (\ref{eq:wcss}) from the main paper), known as Elbow method \cite{Bholowalia_2014}. Figure \ref{fig:elbow} shows the evolution of WCSS as a function of $K$. Although the elbow point is not neatly observed, the shape of the curve suggests that a value of $K$ between 5 and 7 could be a reasonable optimal value.

\begin{figure}[H]
\centering
\includegraphics[width=0.80\linewidth]{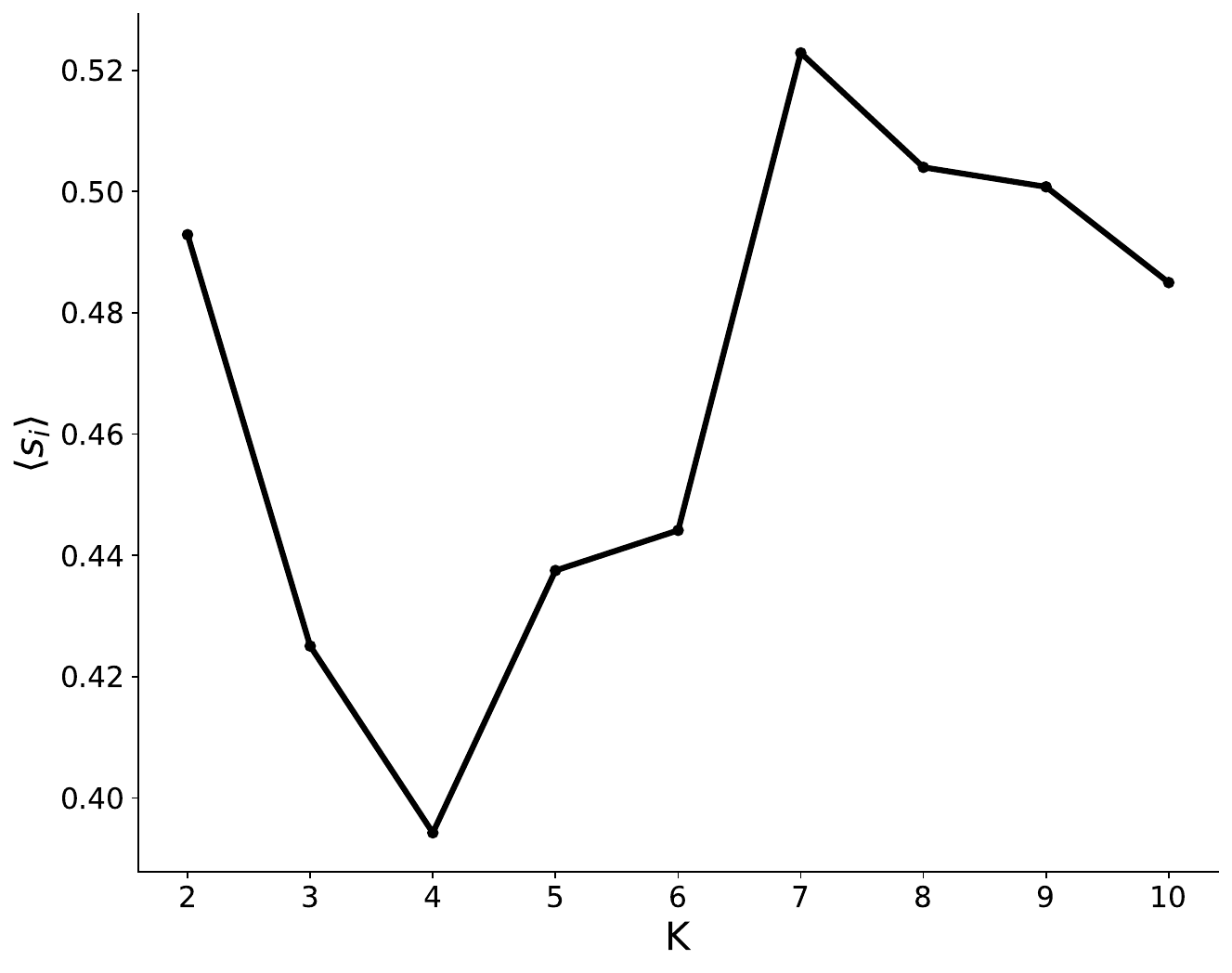}
\caption[Averaged Silhouette value as a function of the number of clusters.]{\textbf{Averaged Silhouette value as a function of the number of clusters.} The Silhouette coefficients $\langle s_{i} \rangle$ represent the separation between clusters. The optimal number of clusters $K$ is chosen as the one that maximizes the averaged Silhouette coefficients over all data points ($K=7$ in this case).}
\label{fig:average_silhouette}
\end{figure}

We also study the Silhouette coefficients \cite{Rousseeuw_1987} ($s_{i}$), which measure the separation between the resulting clusters, i.e., how close each point in a cluster is to the points in the neighboring clusters. Figure \ref{fig:average_silhouette} shows that $K=7$ is the number of clusters that maximize the averaged Silhouette coefficients ($\langle s_{i}\rangle$), and therefore is the optimal $K$.

\begin{figure}[H]
\centering
\includegraphics[page=3, width=0.85\linewidth]{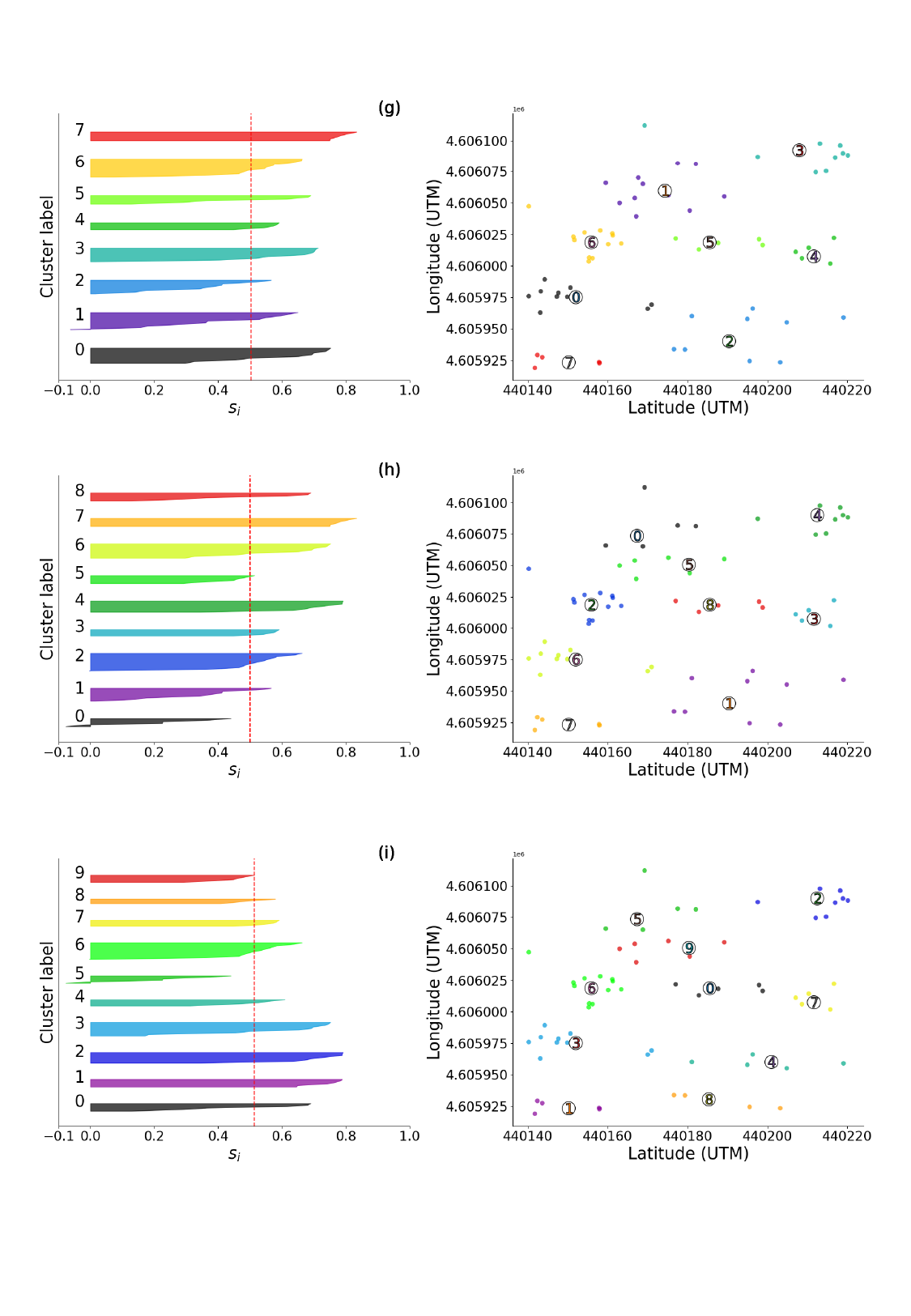}
\vspace{-2cm}
\caption[Individual Silhouette coefficients of each cluster (left panels) and cluster representation (right panels).]{\textbf{Individual Silhouette coefficients of each cluster (left panels) and cluster representation (right panels).} (a) $K=2$. (b) $K=3$. (c) $K=4$. (d) $K=5$. (e) $K=6$. (f) $K=7$. (g) $K=8$. (h) $K=9$. (i) $K=10$. The cluster representation is done using the UTM projection of the latitude and the longitude. The red vertical dotted line is the averaged Silhouette score over all data points $\left( \langle s_{i}\rangle \right)$. We check that $K=7$ is the optimal choice not only because the averaged Silhouette score $\langle s_{i} \rangle$ is maximized but also because most of the coefficients are above (or near) the average value, with practically no negative values.}
\label{fig:silhouette}
\end{figure}

Figure \ref{fig:silhouette} shows 9 panels (from $K=2$ to $K=10$) with the individual Silhouette coefficients and the averaged Silhouette scored over all data points, together with a cluster representation of the data points. We confirm that $K=7$ is the optimal choice for the number of clusters, since the Silhouette coefficients are quite uniform and close to the average value, with practically no negative values. 

\begin{figure*}
\centering
\includegraphics[page=1, width=0.85\linewidth]{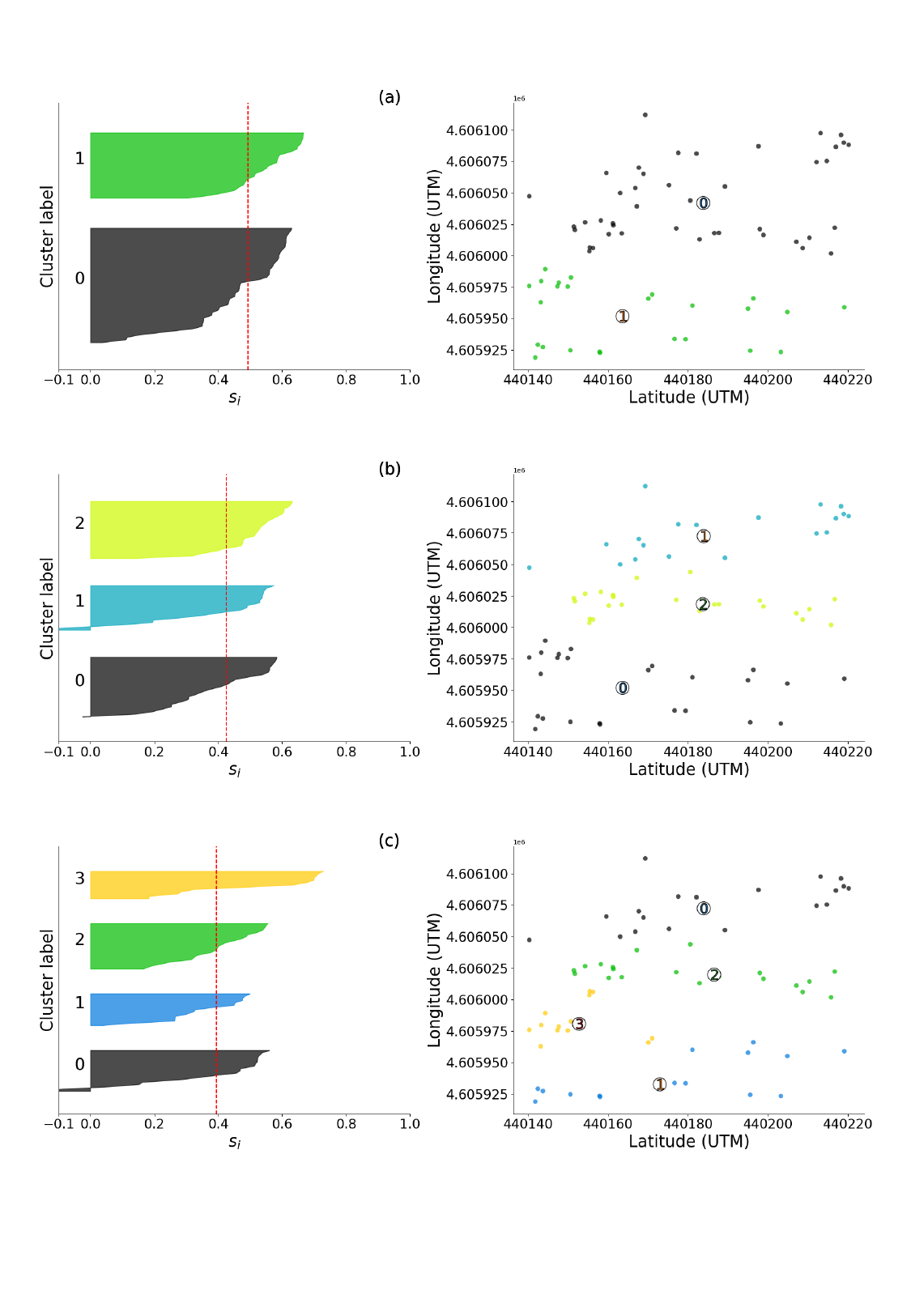}
\end{figure*}

\begin{figure*}
\centering
\includegraphics[page=2, width=0.85\linewidth]{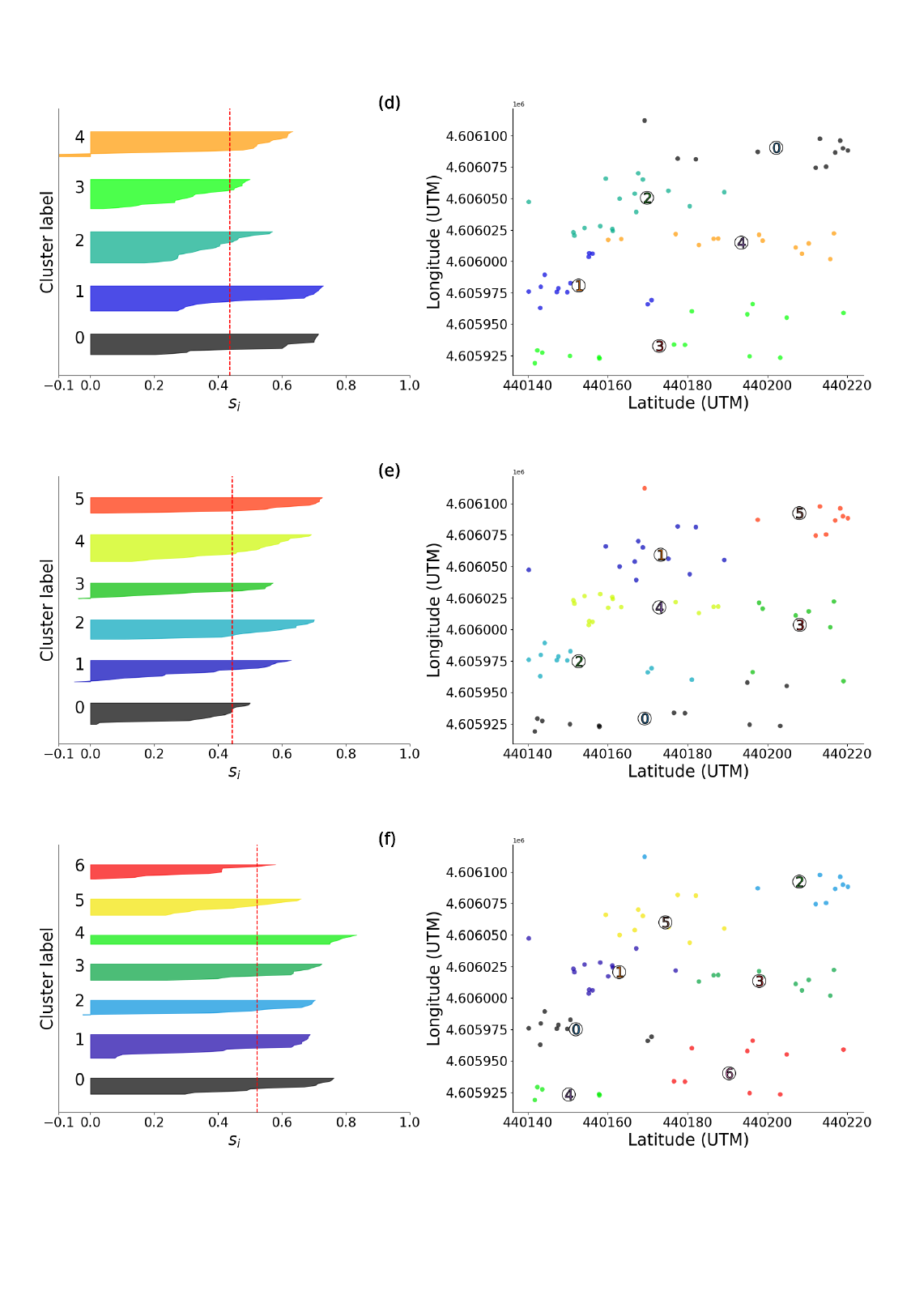}
\end{figure*}

\begin{figure}[H]
\centering
\includegraphics[width=0.80\linewidth]{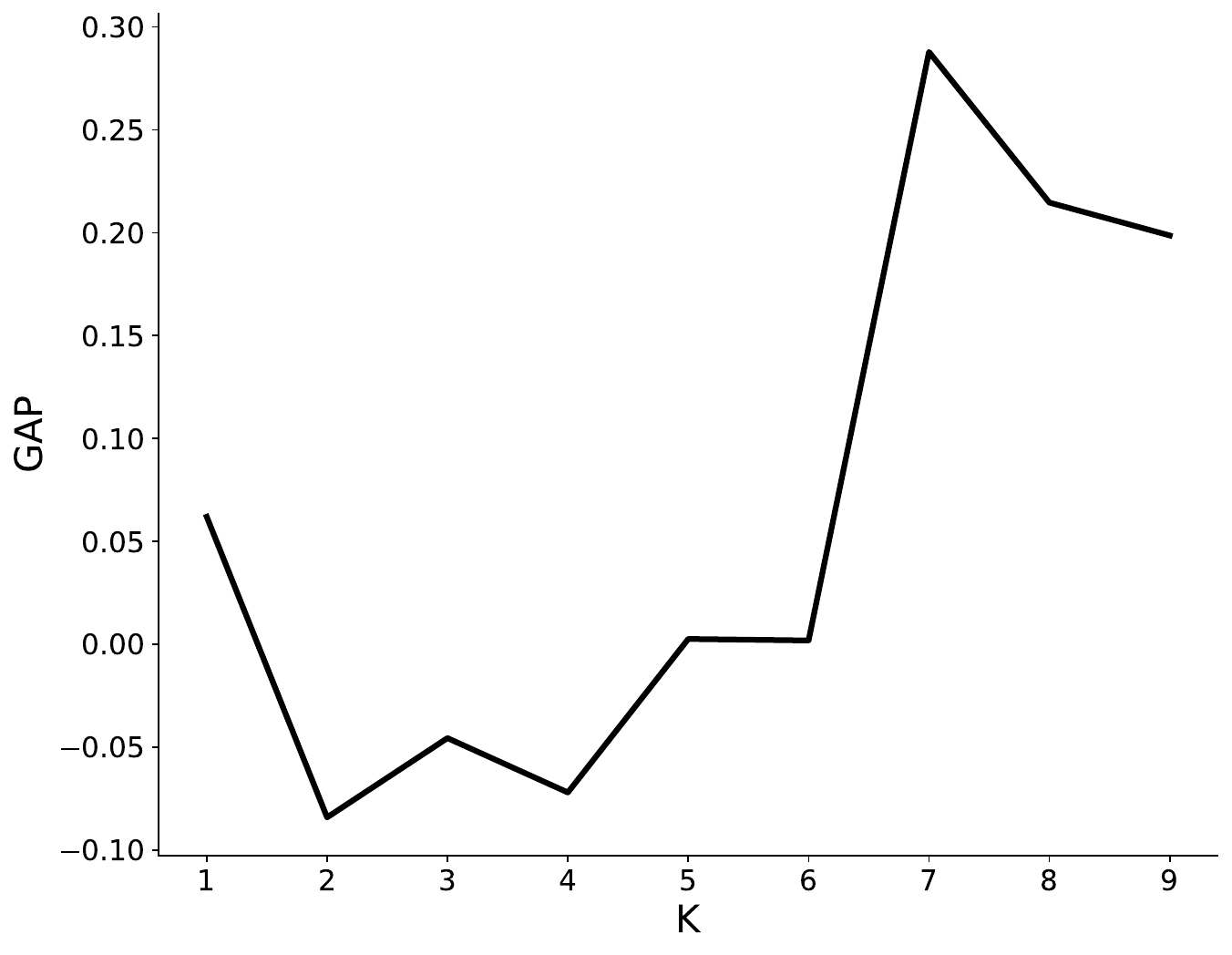}
\caption[GAP statistic as a function of the number of clusters.]{\textbf{GAP statistic as a function of the number of clusters.} The GAP value measures how the clustering structure is far away from the random uniform distribution of data points. The optimal number of clusters $K$ is then chosen to be smallest one that maximizes the GAP($K$), i.e., is within one standard deviation of the GAP($K + 1$). In our case, the optimal number of clusters is $K=7$.}
\label{fig:GAP}
\end{figure}

Finally, using the GAP statistic method \cite{Tibshirani_2001} we also study how the clustering structure is far from a random uniform distribution of data points, comparing the within-cluster sum of squares (WCSS). The optimal number of clusters is the smallest $K$ such the GAP($K$) is maximized, within one standard deviation of the GAP($K+1$). Figure \ref{fig:GAP} shows the evolution of the GAP statistic with $K$. Again, we find $K=7$ to be the number of optimal clusters. 

\subsection{DBSCAN clustering}\label{ss:clusters_actions}
We also explored other algorithms to classify stops in high-density regions. In particular, we applied a well-known density-based algorithm called DBSCAN \cite{Ester_1996}. The main difference is that DBSCAN automatically determines clusters based on the density of data points, whereas $K$-means is a centroid-based clustering (least squares optimization). 
DBSCAN receives two parameters, $min_{points}$ and $\epsilon$, and the number of clusters does not need to be specified before applying the algorithm. It works as follows:

\begin{enumerate}
    \item Computes the distances between each pair of points
    \item For each point (P) it computes how many neighbors it has, using the parameter $\epsilon$ as the maximum distance to consider two points as neighbors. Therefore, considering another point, Q, if $dist(P,Q) <  \epsilon$, then they are neighbors.
    \item The algorithm receives another parameter called $min_{points}$, which is the minimum number of points, neighbors of P, to consider P as a core element (high-density sample).
    \item The clusters are formed iterating over all points, and considering the neighboring points that are core elements.
    \item If another point, say M, is not a core point but is a neighbor of P and can be connected through a chain of core points (for example, going from P to M trough Q), then M is also considered as part of the cluster (border). 
    \item The rest of the points, which are neither boundaries of the cluster nor core points, are classified noise. 
\end{enumerate}

Although it is not necessary to predefine the number of clusters, the algorithm is very sensitive to the choice of both parameters, especially the threshold $\epsilon$. Similar to the Elbow method, to determine the optimal value of $\epsilon$, it is useful to use a $k$-distance plot. The idea is to calculate the average distance of each point from its $k$-nearest neighbors. Next, the $k$-distances are plotted in ascending order. The aim is to determine the ``knee'', which is the threshold where a sharp change occurs along the $k$-distance curve and corresponds to the optimal value of $\epsilon$. For the parameter $min_{points}$, it is very often to use 4 with 2-D data-sets ($min_{points} = 2·D$ for a data-set with D dimensions).

\begin{figure}[H]
\centering
\includegraphics[width=0.99\linewidth]{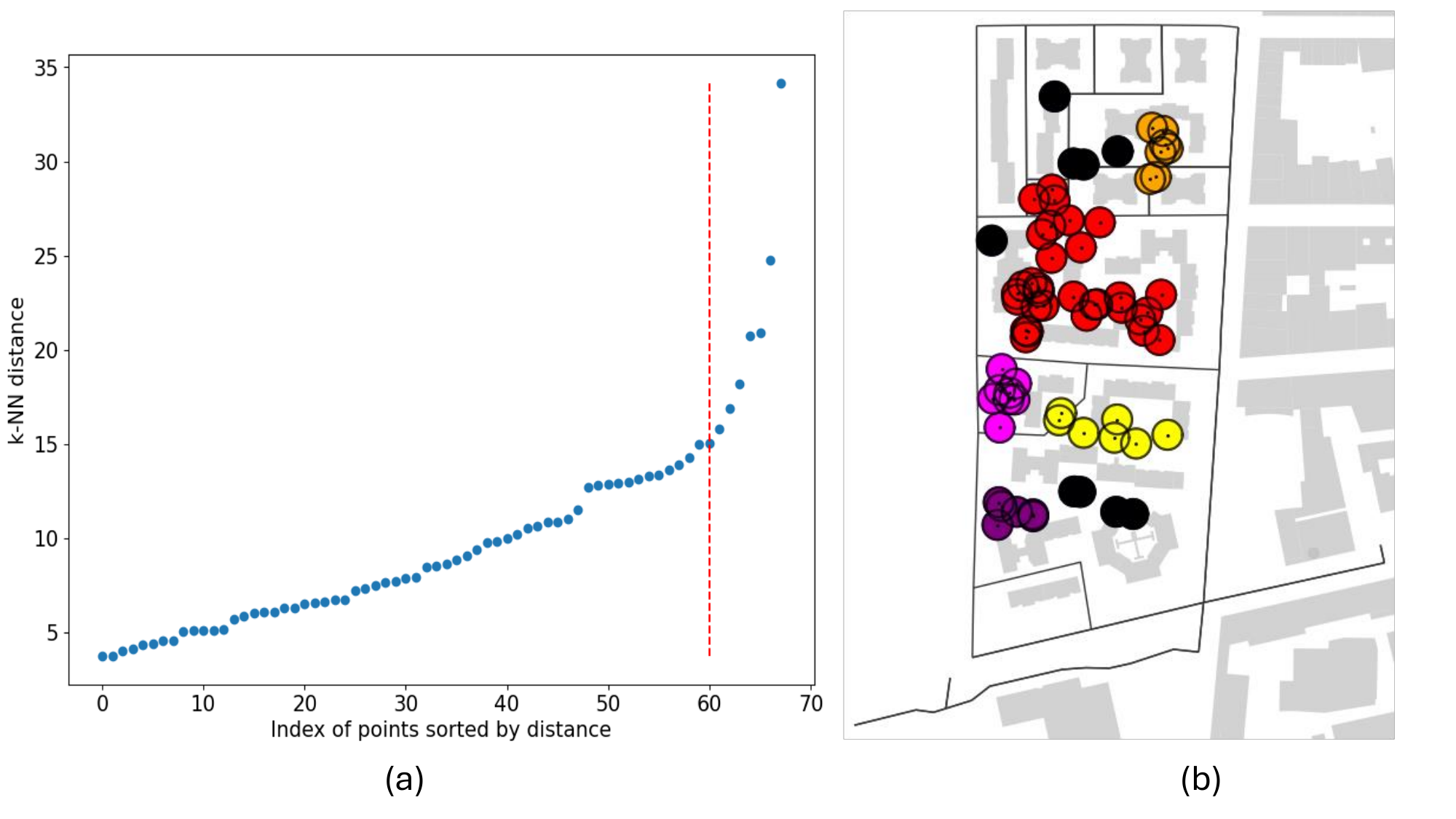}
\caption[DBSCAN density-based algorithm to classify the pausing events.]{\textbf{DBSCAN density-based algorithm to classify the pausing events.} (a) $k$-distance plot to determine the optimal value the parameter $\epsilon$ to run the DBSCAN algorithm. The vertical red dotted line corresponds to the value where the knee of the curve is found. (b) The map representation of the 5 clusters found automatically by the algorithm, with 9 points labelled as noise (in black).}
\label{fig:dbscan}
\end{figure}

Figure \ref{fig:dbscan}a shows the $k$-distance plot with a vertical line corresponding to the index value where a knee is located. The $k$-distance corresponding to that value is the optimal $\epsilon$, which is $\epsilon = 15.07m$. The knee is obtained with a 
method from a library called kneebow (\href{https://github.com/georg-un/kneebow}{https://github.com/georg-un/kneebow}), which uses a Rotor to simply rotate the data, so that the curve looks down and then take the minimum value (elbow). Running the DBSCAN algorithm with $\epsilon = 15.07$ and $min_{data}=4$, 5 clusters are found with 9 points (stops) classified as noise. Figure \ref{fig:dbscan}b shows the map representation of the 5 clusters (black points correspond to noise).

In Table \ref{tab:dbscan_parameters} we show the results (number of clusters and noisy records found by the DBSCAN algorithm) for different combinations of the parameters $\epsilon$ and $\min_{data}$. As can be expected, as $\epsilon$ increases (larger distances for the threshold of neighbors), the number of clusters is reduced, so the algorithm may be capturing more than one high-density region in a single cluster. On contrary, for smaller values of $\epsilon$, the condition to be neighbors is very strict, so there are many clusters and many noisy stops. For a fixed $\epsilon$, changing $\min_{points}$ means changing the minimum number of neighboring points above which we consider a stop as a high density point (core element). Therefore, increasing $\min_{point}$ leads to fewer clusters and more noisy points.

\begin{table}[H]
\centering
\begin{adjustbox}{width=\textwidth}
\begin{tabular}{c|c|c|c|c|c}
\hline \hline
      \diagbox[width=\dimexpr \textwidth/8+2\tabcolsep\relax, height=1cm]{ $\epsilon$ }{$min_{points}$} & 2 & 3 & 4 & 5 & 6\\
      \hline
      10 & \makecell{15 clusters \\ 12 noisy points} & \makecell{9 clusters \\ 24 noisy points} & \makecell{5 clusters \\ 37 noisy points} & \makecell{4 clusters \\ 43 noisy points} & \makecell{2 clusters \\ 53 noisy points} \\ \hline

      12 & \makecell{11 clusters \\ 8 noisy points} & \makecell{7 clusters \\ 16 noisy points} & \makecell{6 clusters \\ 21 noisy points} & \makecell{6 clusters \\ 24 noisy points} & \makecell{5 clusters \\ 33 noisy points} \\
      \hline
      
      15 & \makecell{8 clusters \\ 3 noisy points} & \makecell{5 clusters \\ 9 noisy points} & \makecell{5 clusters \\ 9 noisy points} & \makecell{4 clusters \\ 16 noisy points} & \makecell{5 clusters \\ 17 noisy points} \\
      \hline
      
      18 & \makecell{7 clusters \\ 2 noisy points} & \makecell{5 clusters \\ 6 noisy points} & \makecell{5 clusters \\ 8 noisy points} & \makecell{5 clusters \\ 10 noisy points} & \makecell{4 clusters \\ 15 noisy points} \\
      \hline

      20 & \makecell{4 clusters \\ 2 noisy points} & \makecell{4 clusters \\ 2 noisy points} & \makecell{3 clusters \\ 6 noisy points} & \makecell{4 clusters \\ 8 noisy points} & \makecell{3 clusters \\ 13 noisy points} \\
      \hline

      25 & \makecell{2 clusters \\ 1 noisy point} & \makecell{2 clusters \\ 1 noisy point} & \makecell{2 clusters \\ 1 noisy point} & \makecell{2 clusters \\ 2 noisy points} & \makecell{2 clusters \\ 4 noisy points} \\
      \hline \hline
\end{tabular}
\end{adjustbox}
\caption[Number of clusters and noisy points for different model parameters.]{\textbf{Number of clusters and noisy points for different model parameters.} 
Number of clusters and noisy points obtained for different combinations of DBSCAN model parameters, $\epsilon$ and $min_{points}$. Smaller values of $\epsilon$ than 15 (the optimal found), lead to a huge amount of noisy punts. For greater values, there are few noisy records but the number of clusters is also small (larger distances for the threshold of neighbors imply larger groups). As can be expected, increasing $min_{points}$ leads to few number of clusters.}
\label{tab:dbscan_parameters}
\end{table}

\begin{figure}[H]
\centering
\includegraphics[width=0.99\linewidth]{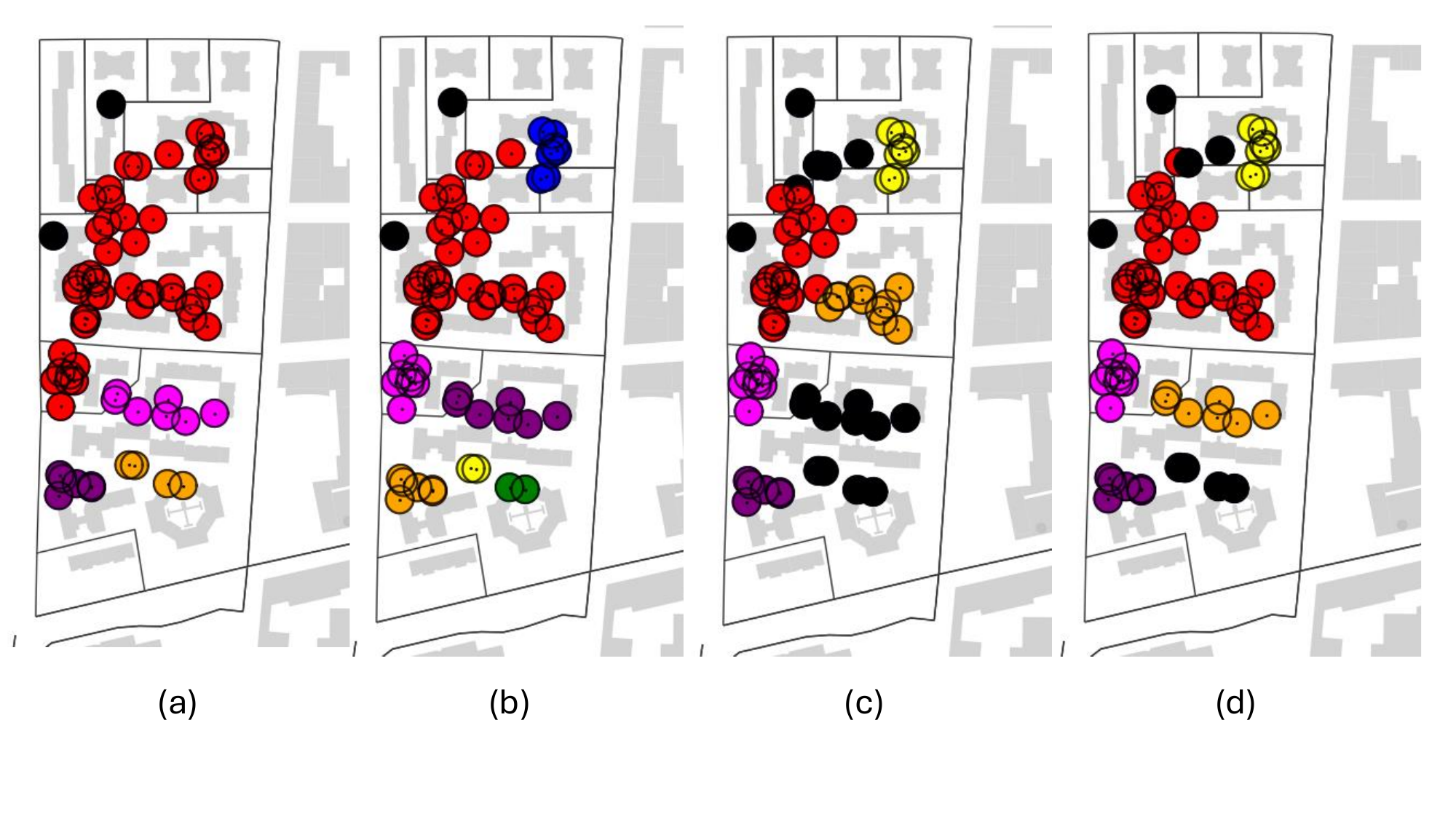}
\caption[DBSCAN clustering results for different model parameters.]{\textbf{DBSCAN clustering results for different model parameters.} (a) For $\epsilon = 20$ and $min_{points}=3$, the algorithm found 4 clusters and 2 noisy points. (b) For $\epsilon = 18$ and $min_{points}=2$, the algorithm found 7 clusters and 2 noisy points. (c) For $\epsilon = 15$ and $min_{points}=6$, the algorithm found 5 clusters and 17 noisy points. (d) For $\epsilon = 18$ and $min_{points}=4$, the algorithm found 5 clusters and 8 noisy points.}
\label{fig:dbscan_param}
\end{figure}

Figure \ref{fig:dbscan_param} shows four clustering representations for different combinations of $\epsilon$ and $min_{points}$. We observe that DBSCAN can not separate the big red cluster as the $K$-means algorithm does. Both Figure \ref{fig:dbscan_param}a and Figure \ref{fig:dbscan_param}b have only 2 noisy records, but in the first case the algorithm is not able to distinguish different high-density regions (big red cluster) and in the other case DBSCAN is overestimating some high-density region (green and yellow clusters). Figure \ref{fig:dbscan_param}c is an example of a very noisy result with a high $min_{points}$ and Figure \ref{fig:dbscan_param}d is similar to Figure \ref{fig:dbscan} using the optimal value of $\epsilon$.

We believe that $K$-means is a better option to capture and distinguish the high-density regions of actions on this high-resolution scale, mainly because DBSCAN it is not able to classify in different clusters the large number of stops in the central part of the neighborhood. 

\section{Description of the clusters}\label{s:clusters_context}
We provide a detailed contextualization of each of the seven identified clusters, focusing on the physical characteristics of their surrounding urban environment, the duration of action stops, and the types of activities performed.

\subsection{Clusters contextualization}

\textbf{AC1 (magenta).} This cluster links two green flowerbed areas encompassing all 10 stops. The actions within this cluster exhibit a playful nature, as detailed in Table \ref{tab:actions_clusters}. The stop durations vary considerably: some last approximately 10–15 minutes, a few are notably brief, and a couple extend to around 30 minutes. Notably, this cluster closely aligns with the P02 area identified in the expert-based mapping (see Fig. \ref{fig:pausing_events}).

\textbf{AC2 (green).} This cluster encompasses a diverse range of actions and is concentrated in a small intermediate area between buildings (see Fig. \ref{fig:pausing_events}), featuring a drinking fountain. It is notable for having the longest individual stop duration, approximately 50 minutes, and the second-highest average stop duration, nearly 18 minutes. The shortest stop duration is also relatively long, and the cluster exhibits the largest standard deviation in stop durations, around 14 minutes (cf. Table \ref{tab:clusters_weighted_kmeans}). The cluster is not identified in the expert-based mapping and just vaguely coincide with C03 area identified  (see Fig. \ref{fig:pausing_events}).

\textbf{AC3 (blue).} This cluster shows similar statistical values to those observed in AC2 (cf. Table \ref{tab:clusters_weighted_kmeans}). Particularly, 8 out of the 10 stops of the cluster are longer than 15 minutes. Actions took place mainly inside green flowerbeds, but some occasional action was performed on the sidewalk and pedestrian crossing. Actions are varied across all topologies but with a clear predominance of celebratory activities (see Table \ref{tab:actions_clusters}). As shown in Fig. \ref{fig:pausing_events}, it follows a very strong overlap with the very prominent C01 area identified in the expert-based mapping.

\textbf{AC4 (orange).} This cluster, comprising 11 actions (9 stops), is situated in an area featuring a playground and the {\it Pla\c ca de la Font}, the neighborhood's sole square and a popular gathering spot for participants post-activity. The average stop duration here is approximately 9 minutes, among the shortest across all clusters. Actions are relatively brief, exhibiting the lowest standard deviation in stop times, suggesting a high degree of consistency in actions duration. Notably, the playground serves as a hub for playful interactions, including ball games (4 out of 11 actions) and social actions such as talking (2 actions) and dancing (2 actions), as detailed in Table \ref{tab:actions_clusters}. However, despite the presence of the square, only one action occurred centrally within it, possibly due to the hard pavement; the majority took place along its periphery, particularly in the surrounding green flowerbeds. Figure \ref{fig:pausing_events} shows a very strong overlap with the very prominent C02 area identified in the expert-based analysis.

\textbf{AC5 (red).} This cluster exhibits the highest density of stops, encompassing 14 stops (20\% of the total 68 stops, corresponding to 24 actions) within a relatively compact area. Despite this high density, average stop duration of this cluster is below the overall average (see Table \ref{tab:clusters_weighted_kmeans}). The area features a pedestrian walkway flanked by numerous green flowerbeds, enhancing its appeal for festive mobility and contributing to the high frequency of actions observed. The area is next to AC2 and it is just been partially identified in the expert-based mapping as part of the P02 area (see Fig. \ref{fig:pausing_events}). Notably, this area is particularly attractive to young people; all seven groups of students from the local school stopped here, with nearly all engaging in two or more actions. Celebratory actions are predominant, accounting for eight of the actions, as detailed in Table \ref{tab:actions_clusters}. Despite the presence of spontaneous and brief interactions, participants also utilized the space for extended periods, engaging in multiple simultaneous activities that emphasized comfort and relaxation. This is evident at the boundary between this cluster and AC2 (green), where two adjacent stops saw groups performing four and five of the six possible actions, with durations of approximately 40 and 50 minutes, respectively. Interestingly, although both stops are in the same location, the second was classified under the AC2 cluster by the $K$-means algorithm. This area seems to be an interesting location to enhance the urban rest capabilities of the neighborhood.

\textbf{AC6 (purple).} This cluster encompasses six stops, resulting in a total of 12 actions with diverse characteristics. Despite the occurrence of simultaneous actions, the stop durations are relatively brief, with a few instances slightly exceeding 15 minutes. The cluster has the fewest stops is neatly compacted within a green flowerbed, which separates two building blocks. The cluster aligns with the P02 area identified in the expert-based analysis (see Fig. \ref{fig:pausing_events}). This region reveals once again the attractiveness of these urban elements (green flowerbeds) when using the public space of the neighborhood. 

\textbf{AC7 (yellow).} This cluster, with 9 stops and 11 actions, partially coincides C04 area identified by the expert-based mapping (see Fig. \ref{fig:pausing_events}). The stop durations are generally brief, with only one exception of a stop exceeding 40 minutes (see Table \ref{tab:clusters_weighted_kmeans}). 
The cluster area is characterized by a building block in the middle. Both sides integrate diverse urban spaces, including green flowerbeds, crosswalks, and two hard-paved public squares (one in each side). This configuration offers a substantial amount of high-quality public space, facilitating a wide range of actions. The cluster accommodates individuals across all age groups, engaging in various activities, highlighting the area's versatility and appeal.

\subsection{Clusters vs Actions}\label{ss:clusters_actionss}
Table \ref{tab:actions_clusters} displays the number of actions performed in each of the weighted $K$-means clusters, depending on the topology of the action. There were 6 social actions to do in form of missions \cite{Larroya_2024}.
\begin{table}[H]
\centering
\begin{adjustbox}{width=\textwidth}
\begin{tabular}{cccccccc}
\hline \hline
& AC1 & AC2 & AC3 & AC4 & AC5 & AC6 & AC7 \\ \hline
J & 5 & 4 & 3 & 4 & 4 & 1 & 3 \\
M & 2 & 2 & 2 & 2 & 3 & 4 & 0 \\
B & 3 & 3 & 2 & 2 & 4 & 1 & 3 \\
X & 0 & 3 & 1 & 2 & 5 & 3 & 3 \\
C+CC & 3 & 6 & 6 & 1 & 8 & 3 & 2 \\ \hline
Total actions & 13 & 18 & 14 & 11 & 24 & 12 & 11 \\ \hline
Stops & 10 & 10 & 10 & 9 & 14 & 6 & 9 \\ 
\hline\hline
\end{tabular}
\end{adjustbox}
\caption[Number of actions of each type in every cluster.]{\textbf{Number of actions of each type in each of the seven activation clusters.} In the left column, the code of each of the actions (J is play, M is lunch, B is dance, X is chat, C is celebration (toast) and CC is celebration (confetti). The last two rows are respectively the total number of actions and the total number of stops in each cluster (which may not coincide as two or more actions could haven been performed at the same stop).}
\label{tab:actions_clusters}
\end{table}
